\documentclass[%
reprint,
amsmath,amssymb,
aps,
pra,
]{revtex4-2}

\usepackage{graphicx}
\usepackage{dcolumn}
\usepackage{bm}

\begin{document}


\title{SU(2) Symmetry of Coherent Photons and Application to Poincar\'e Rotator}

\author{Shinichi Saito}
 \email{shinichi.saito.qt@hitachi.com}
\affiliation{Center for Exploratory Research Laboratory, Research \& Development Group, Hitachi, Ltd. Tokyo 185-8601, Japan.}

\date{\today}

\begin{abstract}
Lie algebra is a hidden mathematical structure behind various quantum systems realised in nature. 
Here, we consider $SU(2)$ wavefunctions for polarisation states of coherent photons emitted from a laser source, and discuss the relationship to spin expectation values with SO(3) symmetry based on isomorphism theorems.
In particular, we found rotated half-wave-plates correspond to mirror reflections in the Poincar\'e sphere, which do not form a subgroup in the projected O(2) plane due to anti-hermitian property.
This could be overcome experimentally by preparing another half-wave-plate to realise a pristine rotator in $SU(2)$, which allows arbitrary rotation angles determined by the physical rotation.
By combining another 2 quarter-wave-plates, we could also construct a genuine phase-shifter, thus, realising passive control over the full Poincar\'e sphere.
\end{abstract}

\maketitle

\section{Introduction}

Marius Sophus Lie introduced the concept of infinitesimal transformations as early as 1870s, which allowed classification and manipulation of complex matrices based on simple sets of Lie brackets, known as commutation relationships by physicists \cite{Stubhaug02,Fulton04,Hall03,Pfeifer03,Dirac30,Georgi99}. 
Lie algebra is especially powerful for applications in quantum mechanics, since the commutation relationships are essential to understanding fundamental properties of elementary particles \cite{Georgi99,Dirac30,Baym69,Sakurai67,Sakurai14}.
One of the most simplest, but yet, non-trivial systems is a quantum 2-level system, described by the special unitary group of 2 dimensions, known as $SU(2)$ \cite{Dirac30,Baym69,Sakurai14}.

These days, $SU(2)$ systems are especially important for applications in quantum computing using qubits \cite{Nielsen00}.
Various qubits are realised by charged-Cooper pairs in superconducting Josephson junctions \cite{Nakamura99,Koch07,Schreier08,Arute19}, ions in optical traps \cite{Bruzewicz19,Pino21}, single photons in silicon photonic circuits \cite{OBrien03,Peruzzo14,Silverstone16,Takeda17}, and single electron spin in silicon transistors  \cite{Lee20,Xue21} for realising Noisy Intermediate-Scale Quantum (NISQ) computing as a near term goal towards the fault-tolerant quantum computing in the long term \cite{Preskill18}.
These qubits are all based on elementary excitations with $SU(2)$ symmetry, and thus, they are fragile against dissipation to environments surrounding microscopic qubits \cite{Caldeira81}. 

On the other hand, polarisation \cite{Max99,Jackson99,Yariv97,Gil16,Goldstein11} is macroscopic manifestation of an spin state of photons with $SU(2)$ symmetry \cite{Jones41,Fano54,Baym69,Sakurai67,Sakurai14}.
The nature of polarisation was successfully discussed by Stokes and Poincar\'e \cite{Stokes51,Poincare92}, even before the discovery of quantum mechanics \cite{Plank00,Einstein05,Bohr13,Dirac28}.
Unlike early days of Stokes and Poincar\'e, today, modern quantum many-body theories are well-established  \cite{Sakurai67,Abrikosov75,Fetter03,Weinberg05,Fox06,Parker05,Altland10} and coherent laser sources are ubiquitously available in experiments \cite{Max99,Jackson99,Yariv97,Gil16,Goldstein11,Hecht17,Pedrotti07}. 
Therefore, we have revisited to understand the nature of polarisation in a coherent state, and found that Stokes parameters, ${\bf S}=(S_1,S_2,S_3)$, are expectation values of spin operators, $\langle \hat{\bf S} \rangle$, and the coherent phases of the $SU(2)$ state were coming from the broken rotational symmetries upon lasing in a vacuum or a waveguide \cite{Saito20a,Saito20b,Saito20c,Saito20d}.
It was also important to recognise that macroscopic number of photons are occupying the same state due to Bose-Einstein condensation, and thus, a simple $SU(2)$ wavefunction is enough to describe the spin state of photons, such that the Poincar\'e sphere is essentially the same as Bloch sphere, except for the fact that the overall factor to represent the magnitude of the total spin is $\hbar N$, where $\hbar$ is the plank constant divided by $2\pi$ and $N$ is the number of photons in the system \cite{Saito20a}.
Our results justify the use of $SU(2)$ wavefunction as a macroscopic wavefunction to describe polarisation, and the impacts of wave-plates or rotators can be understood as quantum mechanical operation to an $SU(2)$ state \cite{Saito20a}.

Here, we consider our $SU(2)$ theory with regard to the relationship to Lie algebra especially for the relationship between the $SU(2)$ state and the observed $\langle \hat{\bf S} \rangle$ with the special orthogonal group of 3-dimensions, $SO(3)$ \cite{Baym69,Sakurai14,Max99,Jackson99,Yariv97,Gil16,Goldstein11,Hecht17,Pedrotti07}.
We discuss how the orbital degrees of freedom are converted to the spin degrees of freedom based on isomorphism theorems in Lie algebra, and confirm the validity of the theory in experiments on polarisation \cite{Max99,Jackson99,Yariv97,Gil16,Goldstein11}.
We also discuss why rotated Half-Wave-Plates (HWPs) behave like pseudo-rotators \cite{Gil16,Goldstein11}, which significantly restrict the use of HWPs for changing the polarisation states.
Based on a simple consideration of Lie algebra, we have solved this issue and confirmed a true rotator could be constructed simply by employing another HWP.
Together with 2 Quarter-Wave-Plates (QWPs), we could also control the amount of the phase-shift simply by the rotation of a HWP.
Consequently, we could construct a passive Poincar\'e controller to realise arbitrary rotations of spin states by mechanical rotations.

\section{Theory}

\begin{figure}[h]
\begin{center}
\includegraphics[width=8cm]{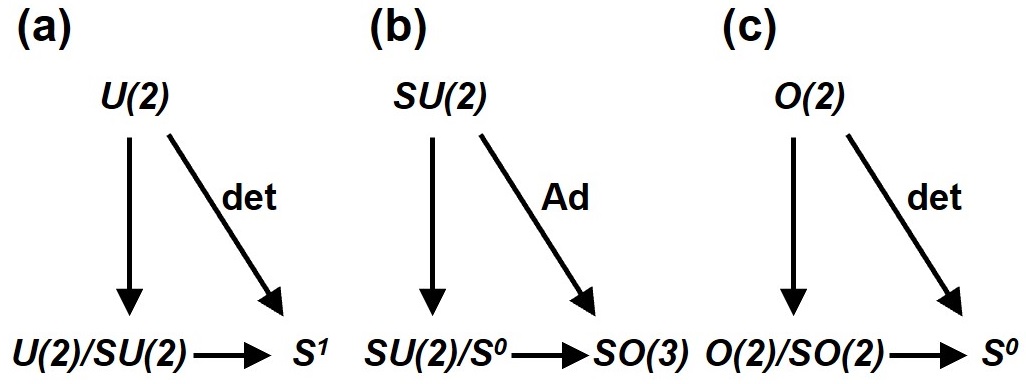}
\caption{
Isomorphic theorems for $U(2)$, $SU(2)$, and $O(2)$.
(a) Isomorphic mapping of $U(2)/SU(2) \cong S^{1}$ induced by a determinant.
(b) Isomorphic mapping of $SU(2)/S^{0} \cong SO(3)$ induced by an adjoint.
(c) Isomorphic mapping of $O(2)/SO(2) \cong S^{0}$ induced by a determinant.
}
\end{center}
\end{figure}

\subsection{$SU(2)$ wavefunction for coherent photons}

A microscopic consideration on spin states of coherent photons was made previously \cite{Saito20a}.
Here, we will review the results \cite{Baym69,Sakurai14,Max99,Jackson99,Yariv97,Gil16,Goldstein11,Hecht17,Pedrotti07} from the perspective of Lie algebra  \cite{Stubhaug02,Fulton04,Hall03,Pfeifer03,Dirac30,Georgi99}.
Our starting point is to accept the principle that coherent photons from a laser are described by a macroscopic wavefunction with 2 degrees of freedom to represent the oscillating electro-magnetic fields perpendicular to each other.
Therefore, the wavefunction contains 2 components, given by 2 complex number ($\mathbb{C}$), which correspond to  2 orbitals for the complex electric fields.
We can choose the basis at our disposal, e.g., by choosing horizontally (H) and vertically (V) linearly polarised, left (L) and right (R) circularly-polarised, or diagonally (D) and anti-diagonally (A) polarised bases \cite{Baym69,Sakurai14,Max99,Jackson99,Yariv97,Gil16,Goldstein11,Hecht17,Pedrotti07}.

The wavefunction must be normalised, such that we have 3 degrees of freedom, given by real number ($\mathbb{R}$).
Topologically, the wavefunction correspond to a point on a surface of a unit sphere in 4-dimensions, $S^{3}=\{ {\bf x} \in  \mathbb{R}^{4} | |{\bf x}|=1 \}$, which is isomorphic to a complex unit sphere in 2-dimension, $S^{1}_{\mathbb{C}}=\{ {\bf x} \in  \mathbb{C}^{2} | |{\bf x}|=1 \}$ \cite{Fulton04,Hall03,Pfeifer03,Georgi99}. In general, we consider a unit sphere in $n$-dimensions with $\mathbb{R}$, $S^{n-1}=\{ {\bf x} \in  \mathbb{R}^{n} | |{\bf x}|=1 \}$, and a complex unit sphere in $n$-dimensions, $S^{n-1}_{\mathbb{C}}=\{ {\bf x} \in  \mathbb{C}^{n} | |{\bf x}|=1 \}$, which is isomorphic to $S^{2n-1}$.
In other words, a quantum mechanical wavefunction corresponds to a point on a surface of a hyper-sphere, describing a state of coherent photons.

We consider a generic transformation of the wavefunction, while we conceive the transformation corresponds to a quantum operation, realised simply by propagation of the electro-magnetic wave into HWPs, QWP, and so on.
The transformation is given by a mapping made by a unitary group of 2-dimension, $U(2)=\{ A \in M(2,  \mathbb{C}) | A^{\dagger} A={\bf 1} \}$, as $U(2)S^{3}\rightarrow S^{3}$, where $M(n,  \mathbb{C})$ is a complex matrix group of $n$-dimensions, $A^{\dagger}$ is an hermitian conjugate (transpose and complex conjugate) of $A$, and ${\bf 1}$ is a unit matrix.
Topologically, this means that a quantum mechanical operation corresponds to a rotation of a state on a surface of a hyper-sphere.
The unitary transformation guarantees the conservation of the norm for the wavefucntion, corresponding to the absence of the loss mechanism during the operation.
In practice, it could be included as an empirical parameter \cite{Max99,Yariv97,Gil16,Goldstein11} for optics, but we will not consider in this work.
The unitary transformation is appropriate to describe systems with time-reversal and space-inversion symmetries.

We consider an surjective mapping of determinant, det, from $U(2)$ to $S^{1} \cong U(1)  \cong \{ {\rm e}^{i \theta} | \theta \in \mathbb{R} \}$.
The sub-group of $U(2)$ with the determinant of unity is $SU(2)=\{ A \in U(2)| {\rm det}(A)=1 \}$, which is the kernel of the mapping of det.
According to the isomorphism theorems in Lie group \cite{Fulton04,Hall03,Pfeifer03,Georgi99}, the projection from $U(2)$ to $U(2)/SU(2)$ induces the isomorphic mapping $U(2)/SU(2) \cong S^{1}$ (Fig. 1(a)).

From a quantum mechanical point of view, above pedagogical mathematics simply means that the wavefunction to describe coherent photons is given by a product of orbital and spin wavefunctions, $U(2) \cong U(1) \times SU(2)$, as
\begin{eqnarray}
\langle z,t
 |\theta, \phi \rangle
&=&
 {\rm e}^{i(kz-\omega t)} 
\left (
  \begin{array}{c}
    {\rm e}^{-i\frac{\phi}{2}}  \cos \left( \frac{\theta}{2} \right)    \\
    {\rm e}^{+i\frac{\phi}{2}}\sin \left( \frac{\theta}{2} \right)  
  \end{array}
\right),
\end{eqnarray}
where $z$ is the direction of propagation, $t$ is time, $\theta$ is the polar angle, $\phi$ is the azimuthal angle in Poincar\'e sphere \cite{Jones41,Fano54,Baym69,Sakurai14,Max99,Jackson99,Yariv97,Gil16,Goldstein11,Hecht17,Pedrotti07,Saito20a}, and we have employed LR-bases \cite{Saito20a}.

In HV-bases, it is given by 
\begin{eqnarray}
\langle z,t 
|\gamma, \delta \rangle
&=&
 {\rm e}^{i(kz-\omega t)} 
\left (
  \begin{array}{c}
    {\rm e}^{-i\delta/2}\cos (\gamma/2)  \\
    {\rm e}^{i\delta/2}\sin (\gamma/2)
  \end{array}
\right) ,
\end{eqnarray}
where $\gamma=2\alpha$ is the azimuthal angle measured from $S_1$ in the Poincar\'e sphere, $\alpha$ is the auxiliary angle, and $\delta$ is the relative phase of the V-state against the H-state \cite{Saito20a}.

\subsection{Lie group of $SU(2)$ for quantum operations}
According to the Lie group theory for $SU(2)$  \cite{Fulton04,Hall03,Pfeifer03,Georgi99,Dirac30,Baym69,Sakurai14,Sakurai67,Jones41,Fano54,Max99,
Jackson99,Yariv97,Gil16,Goldstein11,Saito20a}, the rotation operator in LR-bases along the direction ${\bf \hat{n}}$ ($|\hat{\bf n}|=1$) with the amount of $\delta \phi$ is given by an exponential mapping 
\begin{eqnarray}
\hat{\mathcal{D}}^{\rm LR}({\bf \hat{n}},\delta \phi)
&=&\exp 
\left (
-i 
{\bm \sigma}_{\rm LR} \cdot {\bf \hat{n}}
\left (
\frac{\delta \phi}{2}
\right)
\right), \\
&=& {\bf 1} \cos \left( \frac{\delta \phi}{2} \right) 
-i {\bm \sigma}_{\rm LR} \cdot {\bf \hat{n}} \sin \left( \frac{\delta \phi}{2} \right)
\end{eqnarray}
from Lie algebra using $2 \times 2$ Pauli matrices of ${\bm \sigma}_{\rm LR}=(\sigma_1,\sigma_2,\sigma_3)$, defined as 
\begin{eqnarray}
\sigma_1=
\left(
  \begin{array}{cc}
0 & 1 \\
1 & 0
  \end{array}
\right),
\sigma_2=
\left(
  \begin{array}{cc}
0 & -i \\
i & 0
  \end{array}
\right) , 
\sigma_3=
\left(
  \begin{array}{cc}
1 & 0 \\
0 & -1
  \end{array}
\right). \nonumber \\
\end{eqnarray}
Pauli matrices, $\sigma_i$ ($i=1,2,3$), must satisfy the commutation relationships of Lie algebra ${\mathfrak su(2)}$, which is also known as Lie brackets \cite{Fulton04,Hall03,Pfeifer03,Georgi99} as
\begin{eqnarray}
\left [ \sigma_{i},\sigma_{j} \right ]&=&2 i \epsilon_{ijk}\sigma_{k},
\end{eqnarray}
where $\epsilon_{ijk}$ is the Levi-Civita in 3-dimensions, describing a complete anti-symmetric tensor.
Pauli matrices also satisfy the anti-commutation relationships \cite{Fulton04,Hall03,Pfeifer03,Georgi99} 
\begin{eqnarray}
\left \{\sigma_{i},\sigma_{j}\right \}&=&2\delta_{ij}{\bf 1}.
\end{eqnarray}

For the rotation of $\hat{\mathcal{D}}^{\rm LR}$, we need 3 real parameters, corresponding to ${\bf \hat{n}}$ and $\delta \phi$.
In the original $U(2)$, a general transformation contains 4 real parameters, which includes a phase-shift for the orbital wavefunction of $U(1)$, in addition to $SU(2)$ (Fig. 1(a)).

In HV-bases, we just need to replace ${\bm \sigma}_{\rm LR}$ with ${\bm \sigma}_{\rm HV}=(\sigma_3,\sigma_1,\sigma_2)$, and we obtain\cite{Max99,Jackson99,Yariv97,Gil16,Goldstein11,Saito20a}
\begin{eqnarray}
\hat{\mathcal{D}}^{\rm HV}({\bf \hat{n}},\delta \phi)
&=&\exp 
\left (
-i 
{\bm \sigma}_{\rm HV} \cdot {\bf \hat{n}}
\left (
\frac{\delta \phi}{2}
\right)
\right), \\
&=& {\bf 1} \cos \left( \frac{\delta \phi}{2} \right) 
-i {\bm \sigma}_{\rm HV} \cdot {\bf \hat{n}} \sin \left( \frac{\delta \phi}{2} \right).
\end{eqnarray}
Therefore, the choice of the bases will simply change the axis of rotation. 
For example, the rotation along the $S_1$ axis is performed by $\sigma_3$ in HV-bases \cite{Max99,Jackson99,Yariv97,Gil16,Goldstein11,Saito20a}.

\subsection{Applications of $SU(2)$ theory to optical waveplates and rotators}

An $SU(2)$ theory is powerful to represent operations of optical waveplates and rotators on polarisation states \cite{Jones41,Fano54,Baym69,Sakurai14,Max99,Jackson99,Yariv97,Gil16,Goldstein11,Hecht17,Pedrotti07,Saito20a}.
For example, the impact of the HWP, whose fast-axis/slow-axis (FA/SA) is aligned horizontally/vertically, is represented by setting the $\pi$-rotation as $\delta \phi = \pi$ and the rotation axis along ${\bf \hat{n}_1} = (1,0,0)$.
Then, we obtain $i \hat{\mathcal{D}}^{\rm LR}({\bf \hat{n}}_{1},\pi) \equiv i \hat{\mathcal{D}}^{\rm LR}_{1}(\pi)= \sigma_{1}$ in the LR-bases, or equivalently, it is $i \hat{\mathcal{D}}^{\rm HV}({\bf \hat{n}}_{1},\pi) \equiv i \hat{\mathcal{D}}^{\rm HV}_{1}(\pi)= \sigma_{3}$ in the HV-bases, away from the $U(1)$ phase to describe the overall phase-shift for the propagation of the HWP \cite{Jones41,Fano54,Baym69,Sakurai14,Max99,Jackson99,Yariv97,Gil16,Goldstein11,Hecht17,Pedrotti07,Saito20a}.
The $45^{\circ}$-rotated HWP is also obtained by setting ${\bf \hat{n}_2} = (0,1,0)$, as 
$i \hat{\mathcal{D}}^{\rm LR}({\bf \hat{n}}_{2},\pi) \equiv i \hat{\mathcal{D}}^{\rm LR}_{2}(\pi)= \sigma_{2}$ in the LR-bases and $i \hat{\mathcal{D}}^{\rm HV}({\bf \hat{n}}_{2},\pi) \equiv i \hat{\mathcal{D}}^{\rm HV}_{2}(\pi)= \sigma_{1}$ in the HV-bases \cite{Jones41,Fano54,Baym69,Sakurai14,Max99,Jackson99,Yariv97,Gil16,Goldstein11,Hecht17,Pedrotti07,Saito20a}.
Similarly, for ${\bf \hat{n}}_3=(0,0,1)$, we also obtain the operator of the half-wavelength optical rotator as $i \hat{\mathcal{D}}^{\rm LR}({\bf \hat{n}}_{3},\pi) \equiv i \hat{\mathcal{D}}^{\rm LR}_{3}(\pi)= \sigma_{3}$ in the LR-bases and $i \hat{\mathcal{D}}^{\rm HV}({\bf \hat{n}}_{3},\pi) \equiv i \hat{\mathcal{D}}^{\rm HV}_{3}(\pi)= \sigma_{2}$ in the HV-bases \cite{Jones41,Fano54,Baym69,Sakurai14,Max99,Jackson99,Yariv97,Gil16,Goldstein11,Hecht17,Pedrotti07,Saito20a}.

From mathematical point of view, the origin of the spin rotation was coming from the difference of the phase-shifts in $U(2)$ for orbital components among orthogonal polarisations upon propagation.
For example, HWP gives different phase-shifts due to the difference of the wavelengths along FA and SA, since the refractive indices depend on the directions crystal orientations \cite{Jones41,Fano54,Baym69,Sakurai14,Max99,Jackson99,Yariv97,Gil16,Goldstein11,Hecht17,Pedrotti07,Saito20a}.
In other words, the rotational symmetries are broken in optical waveplates and rotators, and it is effectively equivalent to apply a magnetic field to a magnet, which rotates a spin state.
For a photon, there is no magnetic moment due to the lack charge, but the phase-shift can be precisely controlled by tuning the thickness of waveplates to account for the difference of the rotation upon propagation.
In this sense, optical waveplates and rotators effectively work as a converter to transfer orbital degrees of freedom in $U(2)$ to spin degrees of freedom in $SU(2)$ (Fig. 1 (a)).

\subsection{Mapping from $SU(2)$ to $SO(3)$}
It is well known that $SU(2)$ is isomorphic to $SO(3)$ in Lie group \cite{Fulton04,Hall03,Pfeifer03,Georgi99,Dirac30,Baym69,Sakurai14,Sakurai67,Jones41,Fano54,Max99,
Jackson99,Yariv97,Gil16,Goldstein11,Saito20a}, and we discuss its consequence for coherent photons (Fig. 1 (b)).
For simplicity, we consider LR-bases in this subsection, but the discussion is valid in other bases, simply by replacing axes.
The structure constant of Lie algebra ${\mathfrak su(2)}$ is given by the commutation relationship of Eq. (6), and it is $2i \epsilon_{ijk}$ \cite{Fulton04,Hall03,Pfeifer03,Georgi99,Dirac30,Baym69,Sakurai14,Sakurai67}.
We consider a mapping function of adjoint (Ad) from ${\mathfrak su(2)}$ to ${\mathfrak so(3)}$, 
\begin{eqnarray}
[{\rm Ad}(-i \sigma_i)]_{jk}=2\epsilon_{ijk},
\end{eqnarray}
for components $i,j,k=1,2,3$, respectively (Fig. 1 (b)), which converts the bases from ${\mathfrak su(2)}$ to ${\mathfrak so(3)}$, given by structure constants in ${\mathfrak su(2)}$.
We define the bases in ${\mathfrak so(3)}$ as $[I_i]_{ij}=[{\rm Ad}(-i \sigma_i)]_{jk}/2=\epsilon_{ijk}$, and we obtain 
\begin{eqnarray}
I_1=
\left(
  \begin{array}{ccc}
0 & 0 & 0 \\
0 & 0 & -1 \\
0 & 1 & 0 
  \end{array}
\right),
I_2=
\left(
  \begin{array}{ccc}
0 & 0 & 1 \\
0 & 0 & 0 \\
-1 & 0 & 0 
  \end{array}
\right),
I_3=
\left(
  \begin{array}{ccc}
0 & -1 & 0 \\
1 & 0 & 0 \\
0 & 0 & 0 
  \end{array}
\right), \nonumber \\
\end{eqnarray}
such that the traceless complex $2 \times 2$ matrices, $\sigma_i$, in ${\mathfrak su(2)}$ are replaced with the traceless real $3 \times 3$ matrices of $I_i$, in ${\mathfrak so(3)}$, which satisfy the commutation relationship
\begin{eqnarray}
\left [ I_{i}, I_{j} \right ]&=& \epsilon_{ijk} I_{k}, 
\end{eqnarray}
for ${\bm I}=(I_1,I_2,I_3)$ is angular momentum to generate a rotation \cite{Baym69,Sakurai14}.
The traceless nature of ${\mathfrak su(2)}$ and ${\mathfrak so(3)}$ guarantees the conservation of the norm, such that the number of photons is preserved upon rotational operations to change polarisation states.

The exponential map from Lie algebra ${\mathfrak so(3)}$ to Lie group $SO(3)$ gives a Mueller matrix \cite{Gil16,Goldstein11} 
\begin{eqnarray}
\hat{\mathcal{M}}({\bf \hat{n}},\delta \phi)
&=&\exp 
\left (
{\bm I} \cdot {\bf \hat{n}}
\delta \phi
\right), 
\end{eqnarray}
for coherent photons.
For example, the rotation along the $S_3$ axis is given by ${\bf \hat{n}_3}=(0,0,1)$, and we obtain the Mueller matrix
\begin{eqnarray}
\hat{\mathcal{M}}_3(\delta \phi)
&\equiv&
\hat{\mathcal{M}}({\bf \hat{n}_3},\delta \phi) \\
&=&
\left(
  \begin{array}{ccc}
\cos(\delta \phi) & -\sin(\delta \phi) & 0 \\
\sin(\delta \phi) & \ \ \ \cos(\delta \phi) & 0 \\
0 & 0 & 1 
  \end{array}
\right).
\end{eqnarray}

The commutation relationship of Eq. (6) in ${\mathfrak so(3)}$ is essentially the same as that of Eq. (12) in ${\mathfrak su(2)}$.
However, the mapping from $SU(2)$ to $SO(3)$ is surjective onto-mapping, but it is not injective (Fig. 1(b)).
This could be understood by considering $\delta \phi=2\pi$-rotation in Poincar\'e sphere, which is always $\hat{\mathcal{M}}({\bf \hat{n}},2 \pi) = {\bf 1}$, irrespective to the choice of the rotation axis ${\bf \hat{n}}$, since a unit rotation in a sphere, $S^{2}$, cannot change the position of a point on the sphere after the rotation.
On the other hand, the corresponding rotation in $SU(2)$ changes the signs of wavefunctions $\langle z,t |\theta, \phi \rangle$ of Eq. (1) or $\langle z,t|\gamma, \delta \rangle$ of Eq. (2).
We must account for the factor of 2 difference in rotation angles between $SU(2)$ to $SO(3)$.

This is apparent in the real space image of the wavefunction, since the $SU(2)$ wavefunction is actually describing a complex electric field for orthogonal polarisation components in real space  \cite{Max99,Jackson99,Yariv97,Gil16,Goldstein11,Saito20a}.
Therefore, the $2\pi$-rotation in Poincar\'e sphere corresponds to the $\pi$-rotation in real space, which changes the sign of the electric field, as seen from Eq. (2).
For example, suppose the original input beam is complete horizontally linear polarised state, $|{\rm H} \rangle$.
The application of $2\pi$-rotation could be achieved by 2 successive operations by HWPs, whose FAs are aligned to the same direction.
This will change the input of $|{\rm H} \rangle$ to the output of $-|{\rm H} \rangle$, which is also horizontally polarised state, but has opposite in phase. 
Consequently, the point in the Poincar\'e sphere would not be changed, while the wavefunciton chnges its sign.
This change of the sign could be observed by an interference to the original input beam, which is bypassed from the original input.
In fact, the phase-shift of $\pi$ is ubiquitously employed in a Mach-Zehnder interferometer for high-speed optical switching \cite{Yariv97}.
In reality, of course, we must also consider the $U(1)$ phase-shift, coming form the propagation in HWPs and the difference in optical path lengths, but it can be adjusted.

Mathematically, this is explained by isomorphism theorems (Fig. 1(b)) \cite{Fulton04,Hall03,Pfeifer03,Georgi99}, since the kernel of the adjoint mapping from $SU(2)$ to $SO(3)$ is $\{ {\bf 1},-{\bf 1} \} \cong S^{0} = \{ 1, -1 \}$.
We confirmed this by putting $\delta \phi =2 \pi$ in Eq. (4), which gives the non-trivial change of the sign by $\hat{\mathcal{D}}^{\rm LR}({\bf \hat{n}},2 \pi)=- {\bf 1}$ in $SU(2)$, while we also have a trivial kernel of $\hat{\mathcal{D}}^{\rm LR}({\bf \hat{n}},0)= {\bf 1}$.
On the other hand, in $SO(3)$, both $\hat{\mathcal{M}}({\bf \hat{n}},0)$ and $\hat{\mathcal{M}}({\bf \hat{n}},2 \pi)$ are equivalent to an identity operation, given by a unit $3 \times 3$ matrix of ${\bf 1}$, preserving the point on $S^{2}$.
Therefore, the kernel of $SU(2)$ in the adjoint mapping to $SO(3)$ is indeed $S^{0}$.
Following isomorphism theorems, we obtain $SU(2)/S^{0} \cong SO(3)$.

\subsection{Spin expectation values and Stokes parameters in Poincar\'e sphere}
Now, we have prepared to discuss the application of an $SU(2)$ theory for photonics in more detail.
For coherent photons, we can define the spin operator in $SU(2)$ as
\begin{eqnarray}
\hat{\bf S}
=
\hbar N
\hat{\bm \sigma}
,
\end{eqnarray}
and we use $\hat{\bm \sigma} \rightarrow {\bm \sigma}_{\rm LR}=(\sigma_1,\sigma_2,\sigma_3)$ for LR bases, and $\hat{\bm \sigma} \rightarrow {\bm \sigma}_{\rm HV}=(\sigma_3,\sigma_1,\sigma_2)$ for HV bases \cite{Saito20a}.
By calculating the quantum-mechanical average over $SU(2)$ states, $|\theta, \phi \rangle$ of Eq. (1) or $|\gamma, \delta \rangle$ of Eq. (2),
we obtain
\begin{eqnarray}
\langle \hat{\bf S} \rangle
&=&
\left (
  \begin{array}{c}
    \langle \hat{S}_1 \rangle\\
    \langle \hat{S}_2 \rangle\\
    \langle \hat{S}_3 \rangle
  \end{array}
\right)\\
&=&
\hbar N
\left (
  \begin{array}{c}
    \sin \theta \cos \phi \\
    \sin \theta \sin \phi \\
    \cos \theta 
  \end{array}
\right)  \\
&=&
\hbar N
\left (
  \begin{array}{c}
    \cos \gamma \\
    \sin \gamma \cos \delta \\
    \sin \gamma \sin \delta 
  \end{array}
\right),
\end{eqnarray}
respectively \cite{Saito20a}. 
These average spin values are nothing but Stokes parameters \cite{Saito20a}, such that we confirm ${\bf S} = \langle \hat{\bf S} \rangle$.
We have pointed out that the prefactor of $\hbar N$ is coming from the nature of Bose-Einstein condensation for macroscopic number of photons to occupy the same state with the lowest loss at the onset of lasing  \cite{Saito20a,Saito20b,Saito20c,Saito20d}.

The expectation values of $\langle \hat{\bf S} \rangle$ should not depend on an arbitrary choice of bases, such that we obtain the famous relationships \cite{Max99,Jackson99,Yariv97,Gil16,Goldstein11,Saito20a} for polarisation ellipse as 
\begin{eqnarray}
\tan(2 {\it \Psi})
&=&
\tan(2 \alpha)
\cos\delta \nonumber \\
\sin (2 \chi)
&=&
\sin (2 \alpha) \sin \delta , 
\end{eqnarray}
where the orientation angle is ${\it \Psi}=\phi/2$, and the ellipticity angle is $\chi=\pi/4-\theta/2$.
These are also obtained simply by geometrical considerations of Stokes parameters in Poincar\'e sphere \cite{Max99,Jackson99,Yariv97,Gil16,Goldstein11,Saito20a}.

A general rotation operator in $SU(2)$ \cite{Baym69,Sakurai14} is given by
\begin{eqnarray}
\hat{\mathcal{D}} ({\bf \hat{n}},\delta \phi)
&=&\exp 
\left (
-i 
\hat{\bm \sigma} \cdot {\bf \hat{n}}
\left (
\frac{\delta \phi}{2}
\right)
\right), \\
&=& {\bf 1} \cos \left( \frac{\delta \phi}{2} \right) 
-i  \hat{\bm \sigma} \cdot {\bf \hat{n}} \sin \left( \frac{\delta \phi}{2} \right),
\end{eqnarray}
independent on a choice of bases.
As discussed above, $\hat{\mathcal{D}} ({\bf \hat{n}},\delta \phi)$ acts on the wavefunction in $SU(2)$ to rotate the polarisation state, while the corresponding expectation values become real numbers as spin expectation values of ${\bf S}$, represented in Poincar\'e sphere, which is rotated in $SO(3)$ (Fig. 1(b)).
Both $SU(2)$ and $SO(3)$ form Lie groups \cite{Fulton04,Hall03,Pfeifer03,Georgi99,Dirac30,Baym69,Sakurai14,Sakurai67,Jones41,Fano54,Max99,
Jackson99,Yariv97,Gil16,Goldstein11,Saito20a}, such that rotational transformations are continuously connected to an identity element of ${\bf 1}$ and determinants of group elements are always 1, ensuring the norm conservation.
The adjoint mapping from $SU(2)$ (Eq. (21)) to $SO(3)$ (Eq. (13)) Lie groups is achieved by the corresponding mapping from ${\mathfrak su(2)}$ to ${\mathfrak so(3)}$ Lie algebras as 
\begin{eqnarray}
{\rm Adj} (
-i \hat{\bm \sigma} )
=
{\bm I},
\end{eqnarray}
independent on the choice of the bases.

We can check that a rotation of the polarisation state in $SU(2)$ is actually corresponding to the rotation of the expectation values of spin in $SO(3)$.
Here, we briefly confirm this for optical rotators and phase-shifters in preferred bases.
The optical rotator in LR-bases is given by the rotation along the $S_3$ axis, which is given by
\begin{eqnarray}
\mathcal{R}_{\rm LR}
(\Delta \phi) 
=
\left (
  \begin{array}{cc}
    {\rm e}^{-i\frac{\Delta \phi}{2}} & 0 \\
    0 & {\rm e}^{+i\frac{\Delta \phi}{2}} 
  \end{array}
\right),
\end{eqnarray}
except for the $U(1)$ phase factor (Fig. 1(a)) for the orbital component upon propagation of a quartz rotator or a liquid-crystal rotator, for example as a mean for the chiral rotation \cite{Max99,Jackson99,Yariv97,Gil16,Goldstein11,Saito20a}.
Then, it is straightforward to obtain the output state, $|{\rm output} \rangle $, from the input state, $|{\rm input}\rangle$, as 
\begin{eqnarray}
|{\rm output} \rangle 
&&=\hat{\mathcal{R}}_{\rm LR}
(\Delta \phi) 
|{\rm input}\rangle
\nonumber \\ 
&&=
\left (
\begin{array}{c}
    {\rm e}^{-i\frac{\phi+\Delta \phi}{2}} \cos (\theta/2) \\
    {\rm e}^{+i\frac{\phi+\Delta \phi}{2}}\sin (\theta/2)   
\end{array}
\right),
\end{eqnarray}
which indeed corresponds to rotate the state, $\phi \rightarrow \phi+\Delta \phi$, by a rotator.
In fact, by taking the quantum-mechanical expectation values of the output state, we obtain
\begin{eqnarray}
{\bf S}^{\prime}
\equiv
\langle {\rm output}| {\bf \hat{S}} |{\rm output} \rangle
&=
\hbar N
\left (
  \begin{array}{c}
    \sin \theta \cos(\phi+\Delta \phi)\\
    \sin \theta \sin(\phi+\Delta \phi)\\
    \cos \theta
  \end{array}
\right). \nonumber \\
\end{eqnarray}
The corresponding rotation in $SO(3)$ can also be obtained by Mueller matrix of the rotator for coherent photons \cite{Gil16}, which is actually $\hat{\mathcal{M}}_3(\delta \phi)$ of Eq. (15).
We can immediately recognise that the spin expectation values of Eq. (18) are properly rotated by Eq. (15) to confirm
\begin{eqnarray}
{\bf S}^{\prime}
&=
\hat{\mathcal{M}}_3(\delta \phi)
{\bf S}. 
\end{eqnarray}

For the phase-shifter, on the other hand, it is easier to use HV-bases, and we obtain the phase-shifter operator for an optical waveplate, whose FA is aligned horizontally, as 
\begin{eqnarray}
\Delta_{\rm HV} (\delta_{\rm sf})
&=&
\left (
  \begin{array}{cc}
    {\rm e}^{-i\frac{\delta_{\rm sf}}{2}} & 0 \\
    0 & {\rm e}^{+i\frac{\delta_{\rm sf}}{2}} 
  \end{array}
\right),
\end{eqnarray}
where $\delta_{\rm sf}$ is the expected phase-shift, and we have neglected the overall $U(1)$ phase, as before.
The operator, $\Delta_{\rm HV} (\delta_{\rm sf})$, accounts for the rotation along the $S_1$ axis, as 
\begin{eqnarray}
|{\rm output} \rangle 
&=&\hat{\Delta}_{\rm HV}
(\delta_{fs})
|{\rm input}\rangle
\nonumber \\ &=&
\left (
  \begin{array}{c}
    {\rm e}^{-i\frac{\delta+\delta_{\rm fs}}{2}} \cos \alpha \\
    {\rm e}^{+i\frac{\delta+\delta_{\rm fs}}{2}}\sin \alpha \   
\end{array}
\right),
\end{eqnarray}
which indeed corresponds to a rotation of $\delta \rightarrow \delta+\delta_{\rm fs}$.
Consequently, the spin expectation values become 
\begin{eqnarray}
{\bf S}^{\prime}
\equiv
\langle {\rm output}| {\bf \hat{S}} |{\rm output} \rangle
&=
\hbar N
\left (
  \begin{array}{c}
    \cos (\gamma) \\
    \sin (\gamma) \cos \left( \delta +\delta_{\rm fs}  \right)\\
    \sin (\gamma) \sin \left( \delta +\delta_{\rm fs}  \right)
  \end{array}
\right), \nonumber \\
\end{eqnarray}
which can also be obtained by
\begin{eqnarray}
{\bf S}^{\prime}
&=
\hat{\mathcal{M}}_1(\delta_{\rm fs})
{\bf S},
\end{eqnarray}
where the corresponding Mueller matrix is 
\begin{eqnarray}
\hat{\mathcal{M}}_1(\delta_{\rm fs})
&\equiv&
\hat{\mathcal{M}}({\bf \hat{n}_1},\delta_{\rm fs}) \\
&=&
\left(
  \begin{array}{ccc}
1 & 0 & 0 \\
0 & \cos(\delta_{\rm fs}) & -\sin(\delta_{\rm fs}) \\
0 & \sin(\delta_{\rm fs}) & \ \ \ \cos(\delta_{\rm fs}) 
  \end{array}
\right).
\end{eqnarray}

\subsection{Mirror reflection by rotated half-wavelength phase-shifter}
HWPs, QWPs, and quartz rotators are useful optical components to control polarisation of photons \cite{Max99,
Jackson99,Yariv97,Gil16,Goldstein11}, however, the amounts of rotation are usually fixed, determined by thickness of these plates.
There are several ways to change the amount of rotations \cite{Max99,Jackson99,Yariv97,Gil16,Goldstein11}.
For example, an active control can be made by changing the electric field dynamically upon liquid crystal through transparent electrodes, which is used for applications in a liquid crystal display (LCD) \cite{Yariv97,Gil16,Goldstein11,Hecht17,Pedrotti07}. 
Another method is to rotate a HWP to change the orientation angle of the polarisation ellipse \cite{Max99,Jackson99,Yariv97,Gil16,Goldstein11,Hecht17,Pedrotti07}.
Here, we will revisit the results for impacts on a rotated-HWP and discuss the consequences within a framework of Lie group.

We use LR bases to describe a rotated phase-shifter with the physical rotation angle of ${\it \Delta \Psi}$, and we obtain the operator \cite{Max99,Jackson99,Yariv97,Gil16,Goldstein11,Hecht17,Pedrotti07,Saito20a}
\begin{eqnarray}
&&{\Delta}_{\rm LR}({\it \Delta \phi},\delta_{\rm sf})
=
\mathcal{R}_{\rm LR}({\it \Delta \phi})
{\Delta}_{\rm LR}(\delta_{\rm sf})
\mathcal{R}_{\rm LR}(-{\it \Delta \phi}) \nonumber \\
&&=
\left (
  \begin{array}{cc}
   \cos \left( \frac{\delta_{\rm sf}}{2} \right) & 
   -i {\rm e}^{-i\Delta \phi}  \sin \left( \frac{\delta_{\rm sf}}{2} \right) \\
   -i {\rm e}^{+i\Delta \phi}   \sin \left( \frac{\delta_{\rm sf}}{2} \right) & 
    \cos \left( \frac{\delta_{\rm sf}}{2} \right)
  \end{array}
\right), \nonumber \\
\end{eqnarray}
where the rotator along $\mathcal{R}_{\rm LR}({\it \Delta \phi})=\hat{\mathcal{D}}^{\rm LR}({\bf \hat{n}}_{3},\Delta \phi)$ accounts for the rotation of $\Delta \phi = 2 {\it \Delta \Psi}$ in the Poincar\'e sphere, and ${\Delta}_{\rm LR}(\delta_{\rm sf})=\hat{\mathcal{D}}^{\rm LR}({\bf \hat{n}}_{1},\delta_{\rm sf})$ accounts for the phase-shift of $\delta_{\rm sf}$.

The same result could be obtained by recognising the fact that we need an $SU(2)$ rotation of $\delta_{\rm sf}$ along the tilted direction of ${\bf n}=(\cos ({\it \Delta \phi}),\sin ({\it \Delta \phi}),0)$ in Poincar\'e sphere, and we obtain 
\begin{eqnarray}
&&{\Delta}_{\rm LR}({\it \Delta \phi},\delta_{\rm sf}) \nonumber \\
&&=
\hat{\mathcal{D}}^{\rm LR}((\cos ({\it \Delta \phi}),\sin ({\it \Delta \phi}),0),\delta_{\rm sf}) \nonumber \\
&&=
\cos 
\left(
  \frac{\delta_{\rm sf}}{2}
\right) 
{\bf 1}
-i
\sin 
\left(
  \frac{\delta_{\rm sf}}{2}
\right) 
\left (
  \begin{array}{cc}
    0 &  \exp (-i{\it \Delta \phi})\\
    \exp (+i{\it \Delta \phi}) &   0
  \end{array}
\right), \nonumber \\
\end{eqnarray}

For a HWP, we put $\delta_{\rm sf}=\pi$ to obtain 
\begin{eqnarray}
{\Delta}_{\rm LR}({\it \Delta \phi},\pi) 
=
-i
\left (
  \begin{array}{cc}
    0 &  \exp (-i{\it \Delta \phi})\\
    \exp (+i{\it \Delta \phi}) &   0
  \end{array}
\right),
\end{eqnarray}
which leads the output state of 
\begin{eqnarray}
|\theta^{\prime}, \phi^{\prime} \rangle 
&&=
{\Delta}_{\rm LR}({\it \Delta \phi},\pi) 
|\theta, \phi \rangle 
\nonumber \\ 
&&=
\left (
\begin{array}{c}
    {\rm e}^{-i\frac{2\Delta \phi - \phi}{2}} \cos \left( \frac{\pi-\theta}{2} \right) \\
    {\rm e}^{+i\frac{2\Delta \phi - \phi}{2}} \sin  \left( \frac{\pi-\theta}{2} \right)  
\end{array}
\right).
\end{eqnarray}
Therefore, the impact of a rotated HWP is to change the polar angel, $\theta \rightarrow \theta^{\prime}=\pi - \theta$, and the azimuthal angle, $\phi \rightarrow \phi^{\prime}=2\Delta \phi - \phi$.
This corresponds to the Mueller matrix of 
\begin{eqnarray}
&&\hat{\mathcal{M}}((\cos ({\it \Delta \phi}),\sin ({\it \Delta \phi}),0),\pi) \nonumber \\
&&=
\left(
  \begin{array}{ccc}
\cos(2 {\it \Delta \phi}) & \ \ \ \sin(2 {\it \Delta \phi}) & 0 \\
\sin(2 {\it \Delta \phi}) & -\cos(2 {\it \Delta \phi}) & 0 \\
0 & 0 & -1
  \end{array}
\right),
\end{eqnarray}
which is called as a pseudo rotator \cite{Gil16,Goldstein11}.
The pseudo rotator works as a proper rotator for horizontally/vertically polarised state, since the output polarisation becomes 
\begin{eqnarray}
{\bf S}^{\prime}
=
\left(
  \begin{array}{c}
\pm \cos(4 {\it \Delta \Psi}) \\
\pm \sin(4 {\it \Delta \Psi})  \\
0 
  \end{array}
\right),
\end{eqnarray}
respectively with the rotation angle of 4 times, compared with the physical rotation angle.
However, in general, it does not represent a standard rotation, although ${\Delta}_{\rm LR}({\it \Delta \phi},\pi)$ and $\hat{\mathcal{M}}((\cos ({\it \Delta \phi}),\sin ({\it \Delta \phi}),0),\pi)$ are well-defined operators within $SU(2)$ and $SO(3)$, respectively, with their determinants of 1.

For the $S_3$ component, the pseudo rotation merely changes its sign, such that the left circulation becomes the right circulation, and {\it vice versa}.
Therefore, for the change of the orientation angle, also knowns as the inclination angle to represent the direction of the primary axis of the polarisation ellipse, we consider the projection of $SO(3)$ to its subgroup of $O(2)$ in the $S_1-S_2$ plane (Fig. 2).
Within this plane, the pseudo operation corresponds to the mirror reflection of the original polarisation state (Fig. 2(a)), which is a set of $O^{-}(2)=\{ A \in M(2,  \mathbb{R}) | \det(A)=-1 \}$, given by a mirror matrix \cite{Fulton04,Hall03,Pfeifer03,Georgi99}
\begin{eqnarray}
\hat{\mathcal{M}}_{O_2}(2{\it \Delta \phi}) 
=
\left(
  \begin{array}{cc}
\cos(2 {\it \Delta \phi}) & \ \ \ \sin(2 {\it \Delta \phi}) \\
\sin(2 {\it \Delta \phi}) & -\cos(2 {\it \Delta \phi}) 
  \end{array}
\right)
\end{eqnarray}
in 2-dimensions. 
Interestingly, $O^{-}(2)$ does {not} form a proper sub-group within $O(2)$, since it does not have an identity operator of ${\bf 1}$.
This means that a simple product law as a group like $a\cdot b=c$ for group elements, $a,b$, and $c$, do not necessarily hold.
In particular, we see $\hat{\mathcal{M}}_{O_2}(2{\it \Delta \phi}) \hat{\mathcal{M}}_{O_2}(2{\it \Delta \phi}) ={\bf 1}$, which means the reflection of the reflection brings back to the original state, while the identity is not included in $O^{-}(2)$, ${\bf 1} \notin O^{-}(2)$, such that the mirror reflections are not closed within the set to define the product.

On the other hand, the kernel of $O(2)$ {\it does} form a sub-group of $SO(2)=O^{+}(2)=\{ A \in M(2,  \mathbb{R}) | \det(A)=1 \}$ \cite{Fulton04,Hall03,Pfeifer03,Georgi99}, given by a rotational matrix 
\begin{eqnarray}
\hat{\mathcal{R}}_{O_2}(2{\it \Delta \phi}) 
=
\left(
  \begin{array}{cc}
\cos(2 {\it \Delta \phi}) & -\sin(2 {\it \Delta \phi}) \\
\sin(2 {\it \Delta \phi}) & \ \ \ \cos(2 {\it \Delta \phi}) 
  \end{array}
\right) 
\end{eqnarray}
in 2-dimensions, which is continuously connected to the identity, $\hat{\mathcal{R}}_{O_2}(0) ={\bf 1}$ at $\Delta \phi=0$.
The rotation operators form a group, which is evident from the product of $\hat{\mathcal{R}}_{O_2}(2{\it \Delta \phi}_1) \hat{\mathcal{R}}_{O_2}(2{\it \Delta \phi}_2) =\hat{\mathcal{R}}_{O_2}(2({\it \Delta \phi}_1+{\it \Delta \phi}_2)) $.
According to isomorphism theorems \cite{Fulton04,Hall03,Pfeifer03,Georgi99}, this corresponds to $O(2)/SO(2) \cong S^{0}$.

We understand the pseudo rotator actually works as a mirror reflection within the $S_1-S_2$ plane.
On the other hand, the pseudo rotator is not a complete mirror reflection within the entire Poincar\'e sphere across the mirror plane, defined by a normal vector of $(\sin ({\it \Delta \phi}),-\cos ({\it \Delta \phi}),0)$, which should keep $S_3$ constant.
The pseudo rotator changes the sign of $S_3$, such that the mirror plane for $S_3$ is actually the $S_1-S_2$ plane, whose normal vector is $(0,0,1)$.
As a result, the pseudo rotator could be decomposed of the mirror reflection in the $S_1-S_2$ plane along the direction of $(\cos ({\it \Delta \phi}),\sin ({\it \Delta \phi}),0)$ for $S_1$ and $S_2$ components and another mirror reflection across the $S_1-S_2$ plane for $S_3$.

In order to use the pseudo rotator for realising desired polarisation states, we need to know the input polarisation state {\it a priori} before the application to the rotated-HWP, which limits the application, significantly.
Similar to all other quantum systems, once measurements are taken place, the wavefunction collapses and we cannot recover the original wavefunction completely \cite{Baym69,Sakurai14}.
It is ideal to construct a genuine rotator, which can rotate an expected amount, even without observing the input state.

\begin{figure}[h]
\begin{center}
\includegraphics[width=8cm]{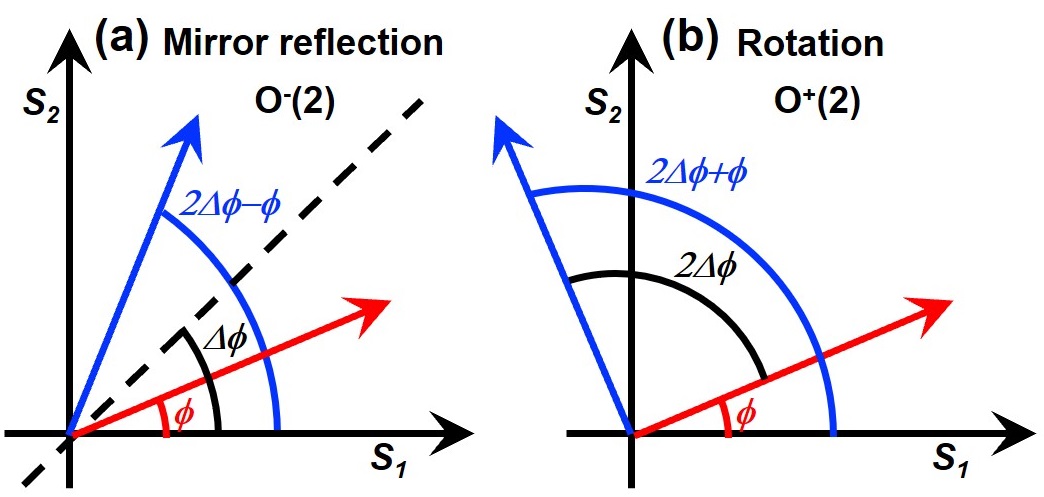}
\caption{
Impacts of $O(2)=O^{-}(2) \cup O^{+}(2)$ operations on polarisation states within the $S_1-S_2$ plane.
The red and blue arrows indicate input and output states, respectively.
(a) Mirror reflection by a pseudo rotator in a set of $O^{-}(2)$.
(b) Genuine rotation in a Lie group of $O^{+}(2)=SO(2)$.}
\end{center}
\end{figure}

\subsection{Genuine rotator by two half-wave-plates}
We can construct a genuine rotator, simply by introducing another HWP, whose FA is aligned horizontally, prior to the application of the pseudo rotator.
In fact, the impact of successive operations of HWPs are calculated as
\begin{eqnarray}
&&{\Delta}_{\rm LR}({\it \Delta \phi},\pi) 
{\Delta}_{\rm LR}(0,\pi)  \nonumber \\
&&=
-
\left (
  \begin{array}{cc}
    0 &  \exp (-i{\it \Delta \phi})\\
    \exp (+i{\it \Delta \phi}) &   0
  \end{array}
\right)
\left (
  \begin{array}{cc}
    0 &  1 \\
    1 &   0
  \end{array}
\right)
, \nonumber \\
&&=
-
\left (
  \begin{array}{cc}
    \exp (-i{\it \Delta \phi}) & 0  \\
    0 &   \exp (+i{\it \Delta \phi})
  \end{array}
\right) \\
&&=
\mathcal{R}_{\rm LR}
(2 {\it \Delta \phi})
=\mathcal{R}_{\rm LR}
(4 {\it \Delta \Psi}), 
\end{eqnarray}
which is indeed a genuine rotator of the angle of $4 {\it \Delta \Psi}$.

The same result can be confirmed in HV-bases as well.
The rotated HWP operator in HV-bases becomes
\begin{eqnarray}
{\Delta}_{\rm HV}({\it \Delta \phi},\pi) 
=
-i
\left (
  \begin{array}{cc}
    \cos({\it \Delta \phi}) &  \ \ \ \sin({\it \Delta \phi})\\
    \sin({\it \Delta \phi}) &   -\cos({\it \Delta \phi})
  \end{array}
\right),
\end{eqnarray}
such that we obtain
\begin{eqnarray}
{\Delta}_{\rm HV}({\it \Delta \phi},\pi) 
{\Delta}_{\rm HV}(0,\pi) 
&=&
-
\left (
  \begin{array}{cc}
    \cos({\it \Delta \phi}) &  -\sin({\it \Delta \phi})\\
    \sin({\it \Delta \phi}) &   \ \ \ \cos({\it \Delta \phi})
  \end{array}
\right), \nonumber \\
&=&
\mathcal{R}_{\rm HV}
(2 {\it \Delta \phi})
=\mathcal{R}_{\rm HV}
(4 {\it \Delta \Psi}), \nonumber \\
\end{eqnarray}
and therefore, we could construct a genuine rotation simply by 2 HWPs, while we must be careful for the amount of rotation of $4 {\it \Delta \Psi}$ (Fig. 2 (b)).
This simply means that the application of another HWP, 
$\hat{\mathcal{D}}^{\rm HV}_{1}(\pi)= -i \sigma_{3}$, converts the pseudo rotator to the genuine rotator in $SU(2)$.
Mathematically, this corresponds to $O^{+}(2)\cong \sigma_3 O^{-}(2)$ within projected $O(2)$.
Consequently, we can control the amount of rotation in Poincar\'e sphere simply by changing the amount of the physical rotation of a HWP in the laboratory.
Having established a proper rotation, it is also straightforward to realise a genuine phase-shifter by inserting 2 QWPs just before and after the genuine rotator, realised by 2 HWPs, since the application of a QWP corresponds to the $\pi/2$-rotation in Poincar\'e sphere \cite{Saito20a}.

\section{Experiments}

\subsection{Experimental set-up}
The experimental set-up is shown in Fig. 3.
We used a frequency-locked distributed-feedback (DFB) laser diode at the wavelength of 1533nm.
The output power was 1.8mW.
The laser is coupled to a single mode fibre (SMF), and the beam is collimated to propagate in a free space, where rotating optical plates are located.
The output beam is collected through a collimator to couple to a SMF.
The polarisation states in SMFs  were controlled by polarisation controllers, which apply stress to induce birefringence in SMFs.
The stress was adjusted prior to experiments to examine the impact of rotating optical plates, inserted within the free space region of the set-up (Fig. 3).
The amount of rotation was physically adjusted by hand with a standard optical rotating element to accommodate wave-plates.
A polarimeter was used to measure the polarisation state.

\begin{figure}[h]
\begin{center}
\includegraphics[width=8cm]{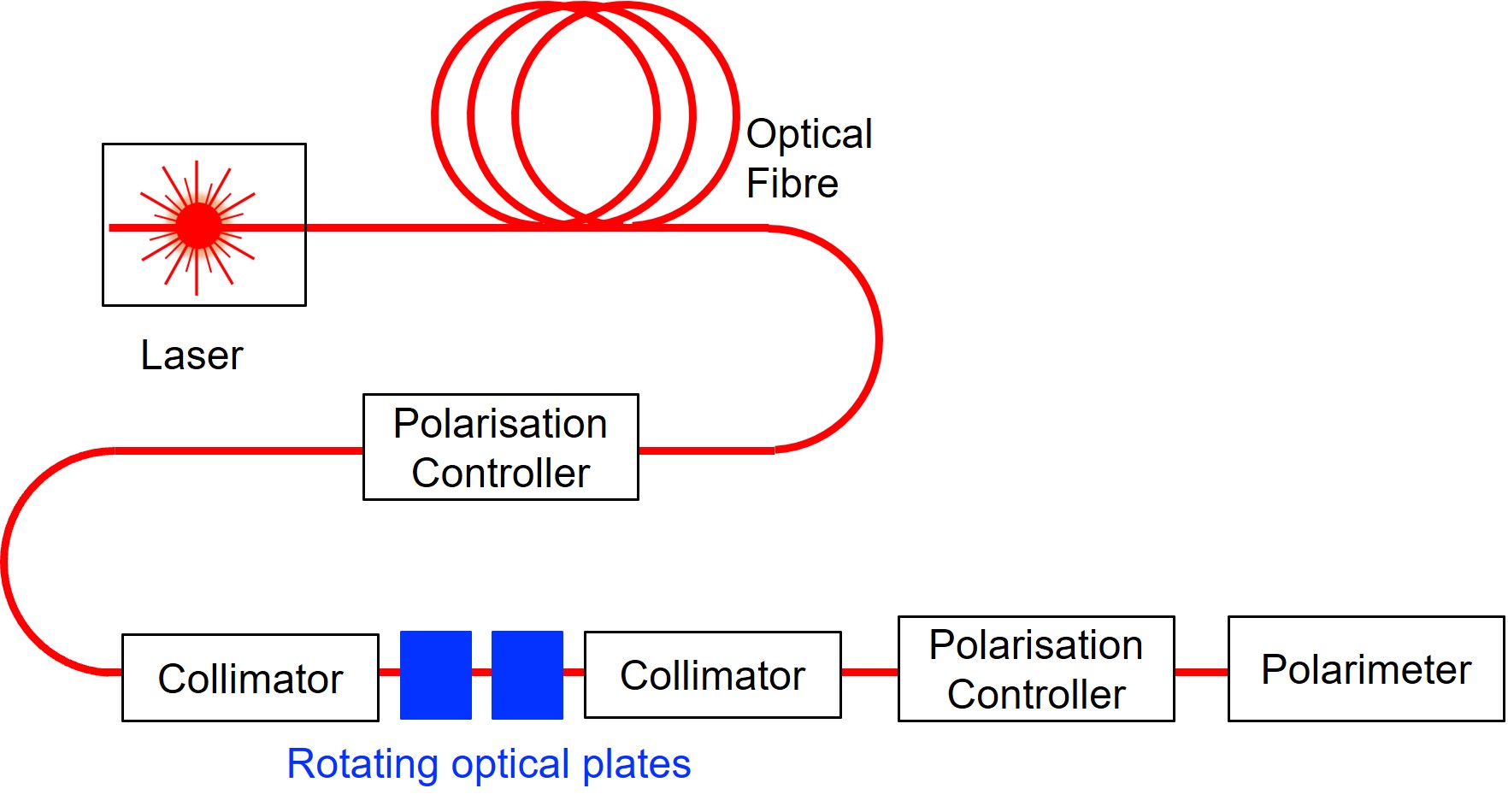}
\caption{
Experimental set-up.
The frequency locked DFB laser diode at the wavelength ($\lambda$) of 1533 nm was coupled to a single mode optical fibre.
Polarisation controllers were used to adjust the polarisation state within the fibres.
The rotating optical plates were inserted in a free space between collimator lenses.  
The output beam was characterised by a polarimeter.
}
\end{center}
\end{figure}

\subsection{Rotated quarter-wave-plates}

\begin{figure}[h]
\begin{center}
\includegraphics[width=8cm]{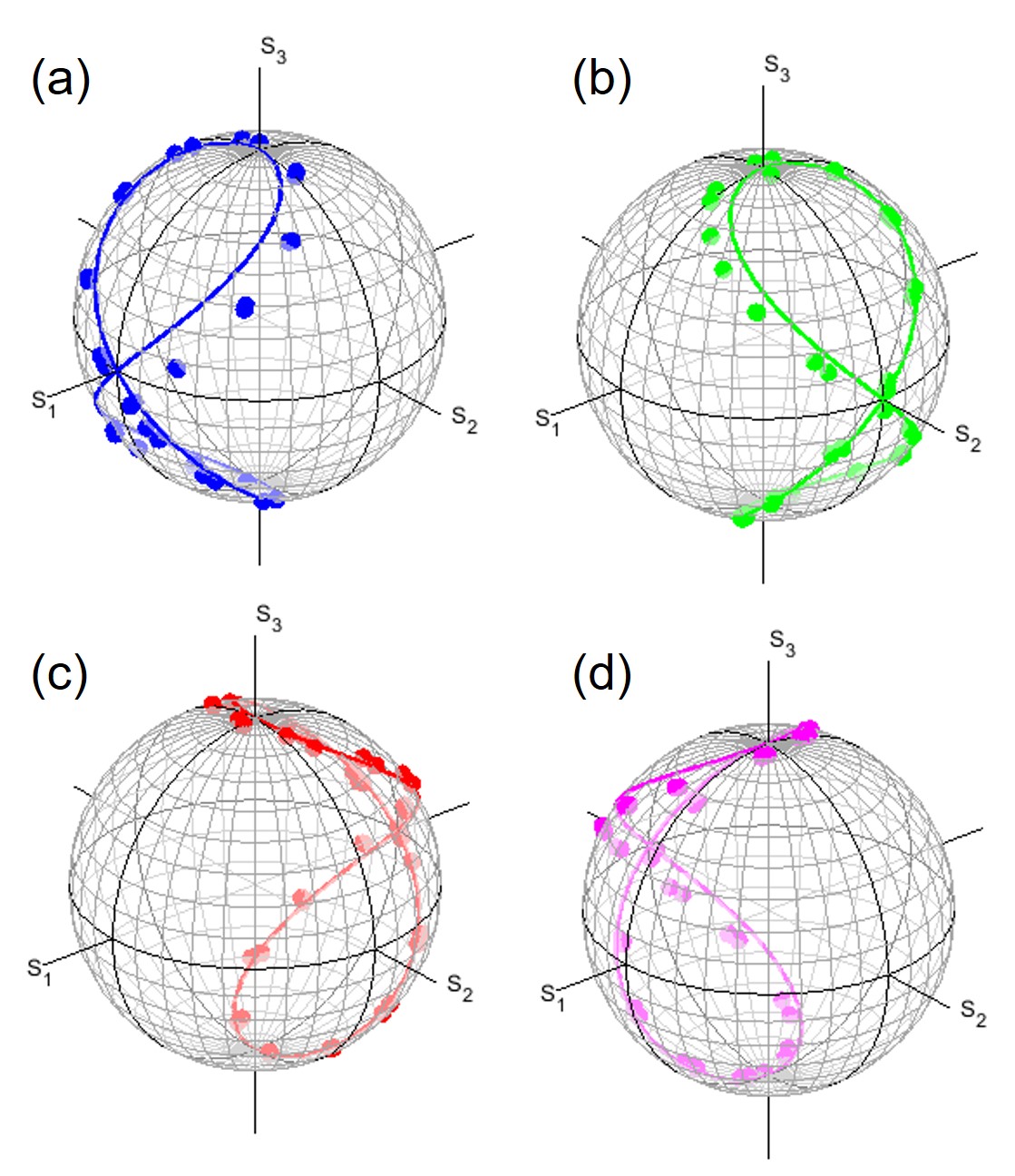}
\caption{
Polarisation states rotated in the Poincar\'e sphere by rotated quarter-wave-plates for inputs of 
(a) horizontally (blue), (b) diagonally (green), (c) vertically (red), and (d) anti-diagonally (magenta) polarised states.
The lines are calculated results and dots are experimental results. 
Circles of latitude (parallels) and circles of longitude (meridians) are shown in every 10$^{\circ}$.
}
\end{center}
\end{figure}

First, we have examined the impacts of rotated QWPs \cite{Max99,Jackson99,Yariv97,Gil16,Goldstein11,Hecht17,Pedrotti07,Saito20a} on polarisation states (Fig. 4).
A QWP, whose FA is aligned horizontally, rotates the diagonally polarised state $|{\rm D} \rangle$ to the left circularly polarised state $|{\rm L} \rangle$ \cite{Max99,Jackson99,Yariv97,Gil16,Goldstein11,Hecht17,Pedrotti07,Saito20a}, while it preserves the horizontally polarised state, $|{\rm H} \rangle$ , and vertically polarised state, $|{\rm V} \rangle$, since it corresponds to rotate the state for 90$^{\circ}$ along the $S_1$ axis.
For the definition on the rotation, we followed the notation of \cite{Jackson99,Saito20a} to see the locus of the electric field, seen from a detector side in the right-handed coordinate.
By changing the physical rotation angle, ${\it \Delta \Psi}$, of the QWP, the polarisation state would be continuously rotated with the maximum change of $\pm 90^{\circ}$.
Theoretical expectation values could be calculated by the $SU(2)$ theory \cite{Max99,Jackson99,Yariv97,Gil16,Goldstein11,Hecht17,Pedrotti07,Saito20a}.
For example, if the input is the horizontally polarised state, the spin expectation value ${\bf S}^{\prime}$ of the output state becomes
\begin{eqnarray}
{\bf S}^{\prime}
&=&
\hbar N
\left (
  \begin{array}{c}
     \cos^2 ({\it \Delta \phi}) \\
     \sin ({\it \Delta \phi}) \cos ({\it \Delta \phi}) \\
     - \sin ({\it \Delta \phi}) \\
  \end{array}
\right),
\end{eqnarray}
where the amount of rotation angle in the Poincar\'e sphere is defined to be ${\it \Delta \phi}=2{\it \Delta \Psi}$, as before.

The comparison between experiments and theoretical calculations are shown in Fig. 4. Our optical module for the physical rotation of a wave-plate has the accuracy of $\pm 5.5^{\circ}$, which dominates the deviation from theoretical calculations.
We also expect the deviation of the retardance from $\lambda/4$ with the amount of $0.006 \lambda$, which corresponds to the additional uncertainty of $\pm 2.2^{\circ}$.
The situation could be worth, since the amount of the rotation in the Poincar\'e sphere could be twice of that in the real space, as seen from Eq. (46).
In fact, the maximum deviations of the order of $\pm 10^{\circ}$ were found.
Nevertheless, the overall trends of experimental data are consistent with the theoretical expectations.

We have also examined the impacts of rotated HWPs, and confirmed expected behaviours on the changes of the polarisation states as a pseudo rotator.
In particular, it did not change the polarisation states for the inputs of $|{\rm H} \rangle$ and $|{\rm V} \rangle$, if we set the FA of the HWP to the horizontal direction, while the inputs of $|{\rm D} \rangle$ and $|{\rm A} \rangle$ are converted to $|{\rm A} \rangle$ and $|{\rm D} \rangle$, respectively, for the same set-up.
The changes of polarisation states upon the rotations of HWPs are consistent with theoretical expectations as pseudo rotators.

\subsection{Genuine rotator by 2 half-wave-plates}

\begin{figure}[h]
\begin{center}
\includegraphics[width=8cm]{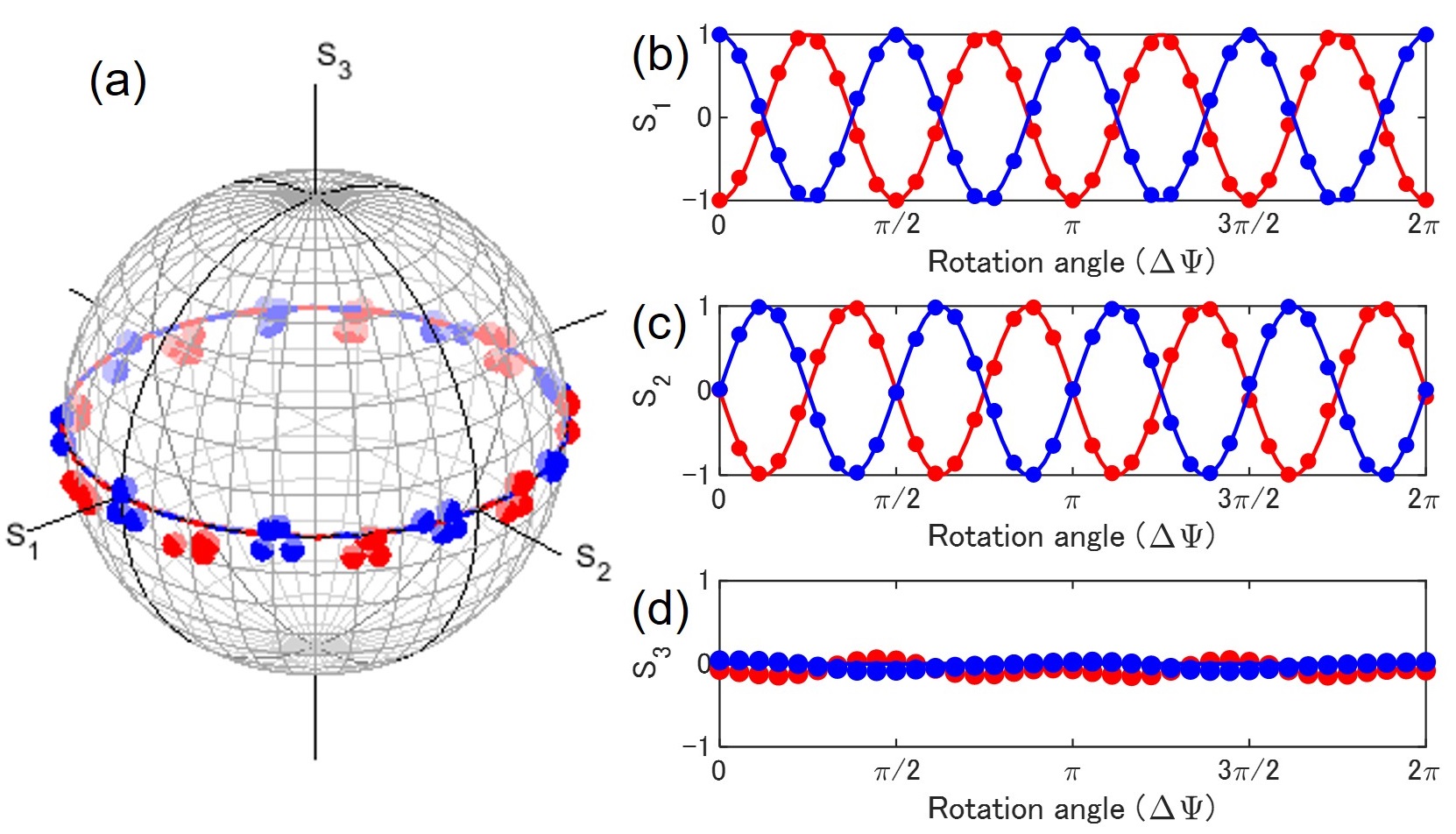}
\caption{
Rotator operation by rotated half-wave-plates for inputs of 
horizontally (blue) and vertically (red) polarised states.
One plate was rotated, while another one was fixed.
(a) Trajectories of polarisation states in the Poincar\'e sphere.
(b) $S_1$, (c) $S_2$, and $S_3$ changed upon the physical rotation (${\it \Delta \Psi}$) of the half-wave-plate.
}
\end{center}
\end{figure}

\begin{figure}[h]
\begin{center}
\includegraphics[width=8cm]{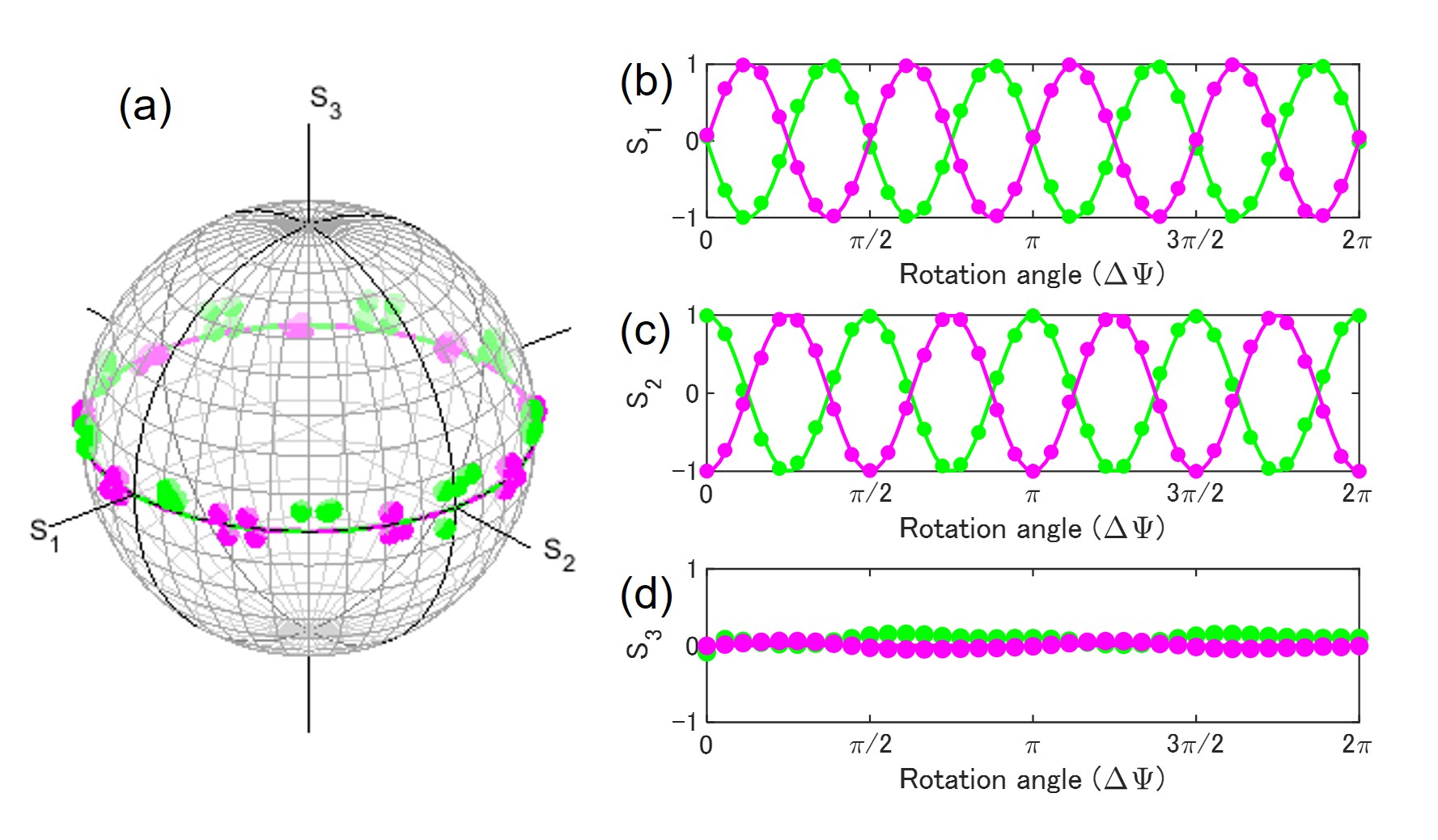}
\caption{
Rotator operation by rotated half-wave-plates for inputs of 
diagonally (green) and anti-diagonally (magenta) polarised states.
One plate was rotated, while another one was fixed.
(a) Trajectories of polarisation states in the Poincar\'e sphere.
(b) $S_1$, (c) $S_2$, and $S_3$ changed upon the physical rotation (${\it \Delta \Psi}$) of the half-wave-plate.
}
\end{center}
\end{figure}

Next, we have set 2 half-wave-plates, one fixed to align the FA horizontally and the other one to allow rotations, as discussed above to realise a genuine rotator.
The experimental results and theoretical comparisons are shown in Figs. 5 and 6.
We see that the polarisation states are rotating 4 times upon the physical 1 rotation of the HWP, as discussed theoretically.
The important evidence as a genuine rotator was confirmed at ${\it \Delta \Psi}=0$, which conserved the polarisation states, such that the input polarisations were preserved, regardless of the inputs.
For Fig. 5, we used $|{\rm H} \rangle$ and $|{\rm V} \rangle$ as inputs, and we observed essentially the same results with those of a pseudo rotator, since the $\pi$-rotation along $S_1$ did not affect $|{\rm H} \rangle$ and $|{\rm V} \rangle$.
On the other hand, $|{\rm D} \rangle$ and $|{\rm A} \rangle$ were reversed by a pseudo rotator (not shown) at ${\it \Delta \Psi}=0$.
As shown in Fig. 6, we confirmed that a genuine rotator did not affect the inputs of $|{\rm D} \rangle$ and $|{\rm A} \rangle$ at ${\it \Delta \Psi}=0$.
This is essentially coming from $(-i\sigma_3)^2=-{\bf 1}$ in HV-bases, whose sign does not affect ${\bf S}$ in $SO(3)$.
Therefore, the behaviours of Fig. 6 by a genuine rotator for $|{\rm D} \rangle$ and $|{\rm A} \rangle$ were different in a pseudo rotator.

In the genuine rotator, we can control the amount of rotation in the Poincar\'e sphere solely by controlling the physical amount of rotation irrespective of the input state, which was remarkably different from the behaviour of a pseudo rotator.
Both genuine and pseudo rotators did not affect the $S_3$ component  such that the inputs of linearly polarised state were still linearly polarised states upon the propagation of these rotators.

\subsection{Comparison between genuine and pseudo rotators}

On the other hand, if the inputs contain the $S_3$ component, the difference of the impacts between genuine and pseudo rotators was outstanding.
In Fig. 7, we show the comparison of output states controlled by these rotators for the same input of the polarisation state at $(S_1,S_2,S_3)=(0.71,0,0.71)$.
As expected for a pseudo rotator, we confirmed the sign of the $S_3$ component was changed \cite{Gil16,Goldstein11,Saito20a}, which means the direction of oscillation in the polarisation ellipse was reversed to be the clockwise rotation from the anti-clockwise rotation.
This is inevitable, since the pseudo rotation is coming from a $\pi$-rotation along some rotation axis in the $S_1-S_2$ plane.
Therefore, $S_3$ must change its sign upon the rotation.
As a result, the pseudo rotator cannot recover the original input state, no matter how much we rotate the HWP.
Mathematically, this was from the fact that pseudo rotators do not form a group, and $O^{-}(2)$ does not include the identity operation.

On the other hand, a genuine rotator is composed of 2 rotations, one is a $\pi$-rotation along the $S_1$ axis and the other is a successive $\pi$-rotation along some rotation axis in the $S_1-S_2$ plane.
Therefore, $S_3$ is kept constant upon the total $2\pi$-rotation, while $S_1$ and $S_2$ components are rotated along the $S_3$ axis.
Consequently, the genuine rotator change the polarisation state within the plane, which includes the original point for the input polarisation state.
Ultimately, this is the evidence that the genuine rotators indeed form a subgroup of $SO(2)$, which must include the identity operator of ${\bf 1}$ to maintain the original state.

\begin{figure}[h]
\begin{center}
\includegraphics[width=4cm]{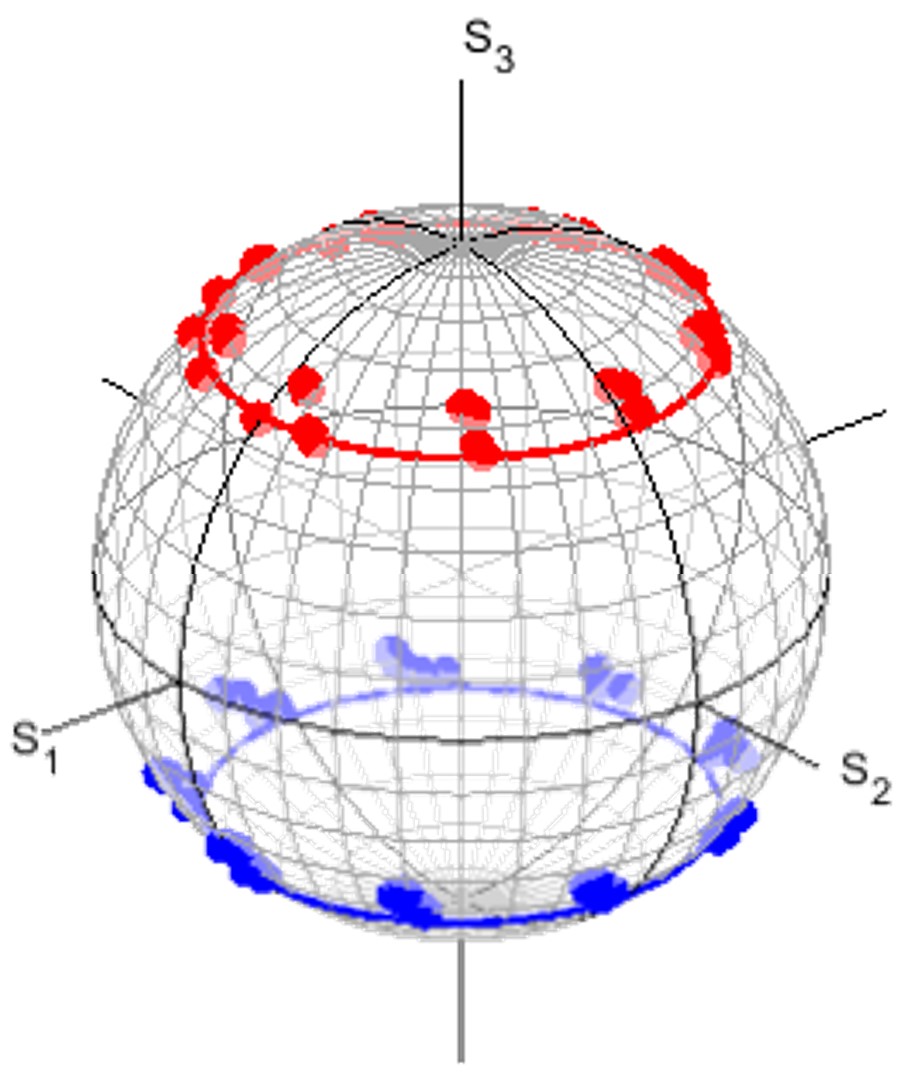}
\caption{
Comparison of genuine (red) and pseudo (blue) rotators on polarisation states in the Poincar\'e sphere.
The polarisation state of the input was located at $(S_1,S_2,S_3)=(0.71,0,0.71)$.
The pseudo rotator changed the sign of $S_3$, such that the chirality is reversed.
The genuine rotator preserved the value of $S_3$, such that the rotation plane includes the original point.
}
\end{center}
\end{figure}

In order to confirm the further evidence that a genuine rotator is different from a pseudo rotator, we consider 2 successive operations of these rotators.
We prepared 2 rotators and the input beam was successively passing through these operators, and we observed the output polarisation state.

For genuine rotators, we expect 
\begin{eqnarray}
\mathcal{R}(4 {\it \Delta \Psi})
\mathcal{R}(4 {\it \Delta \Psi})
=
\mathcal{R}(8 {\it \Delta \Psi}),
\end{eqnarray}
which means that genuine rotation form a group, such that 2 successive operations could be considered to be equivalent to 1 operation of the added rotation angle.
In order to confirm this, we needed to prepare 4 HWPs.
FA of the first one was aligned horizontally, the second one was rotated for ${\it \Delta \Psi}$, and FA of the third one was aligned horizontally, and the forth one was rotated for ${\it \Delta \Psi}$.
The experimental results are shown in Fig. 8.
We confirmed 8 rotations of the polarisation states in the Poincar\'e sphere.
We admit the noticeable fluctuations of experimental data due to physical rotations of 2 HWPs, but they were well below the potential maximum deviations of $\sim \pm 44^{\circ}$ due to 8 times rotations, compared with the physical rotation.

\begin{figure}[h]
\begin{center}
\includegraphics[width=8cm]{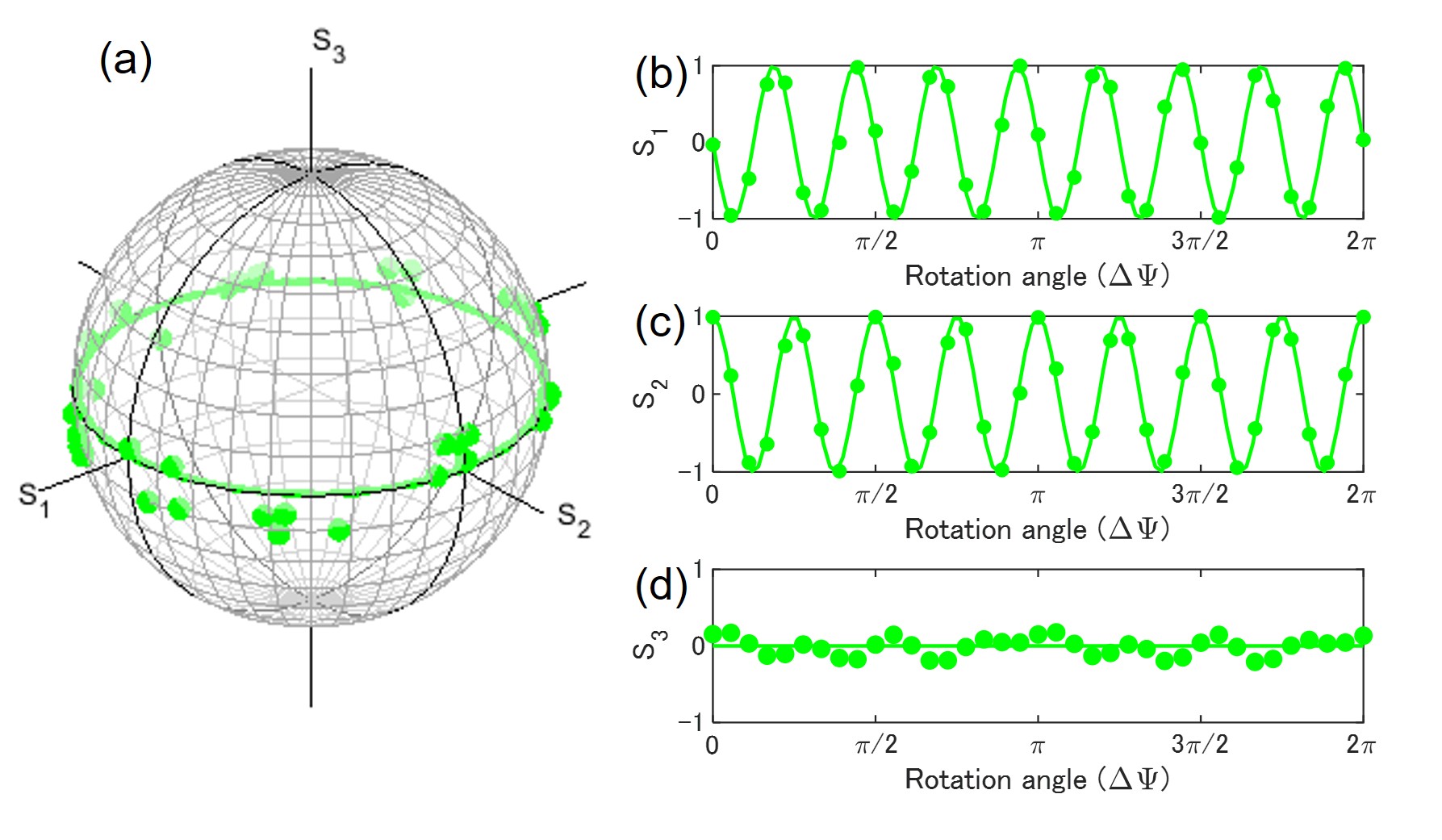}
\caption{
Successive operations of genuine rotators in the Poincar\'e sphere. 
The input state was diagonally polarised.
2 rotators rotated twice of the rotation for 1 rotator.
8 rotations are realised by physical 1 rotation for each rotator. 
}
\end{center}
\end{figure}

On the other hand, 2 successive operations of pseudo rotators should bring the input state back, because a mirror reflection works as an inverse of itself, as
\begin{eqnarray}
\mathcal{M}(4 {\it \Delta \Psi})
\mathcal{M}(4 {\it \Delta \Psi})
=
{\bf 1}, 
\end{eqnarray}
which immediately leads
\begin{eqnarray}
\mathcal{M}^{-1}(4 {\it \Delta \Psi})
=
\mathcal{M}(4 {\it \Delta \Psi})
.
\end{eqnarray}
Therefore, 2 rotators of the same rotation angle cannot change the polarisation state.
In order to confirm this, we needed 2 HWPs, which were rotated at the same angle.
As shown in Fig. 9, we confirmed the polarisation states of output beams were not significantly affected.
Therefore, pseudo rotators are essentially made of mirror reflections, such that 2 successive operations cannot change the input state.

\begin{figure}[h]
\begin{center}
\includegraphics[width=8cm]{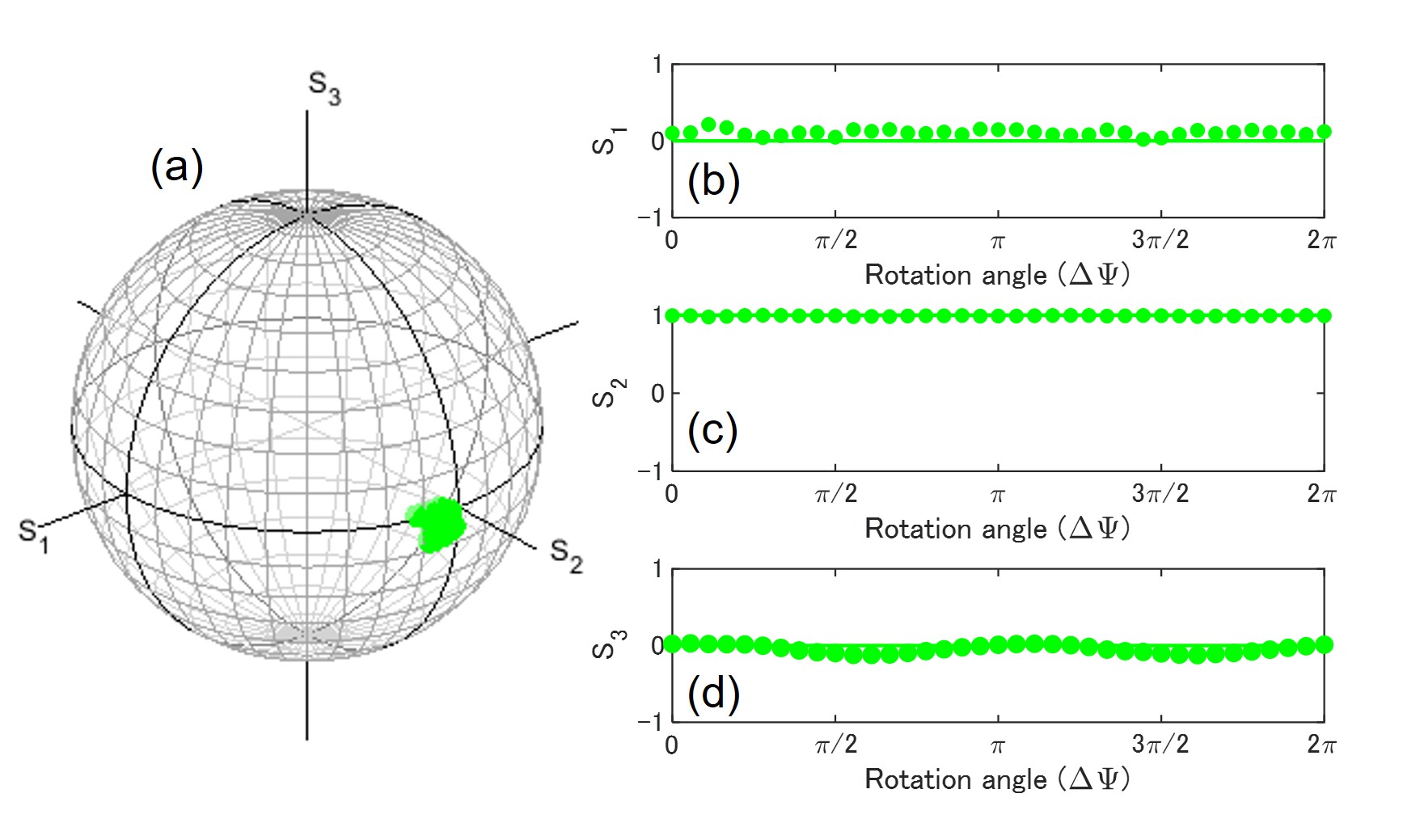}
\caption{
Successive operations of pseudo rotators in the Poincar\'e sphere. 
The input state was diagonally polarised.
This corresponds to 2 mirror reflections, which cannot change the polarisation state.
}
\end{center}
\end{figure}

\subsection{Genuine phase-shifter realised by half-wave and quarter-wave plates}

Now, we could establish how to make a genuine rotator solely by 2 HWPs.
Next, we will show how to construct a genuine phase-shifter, whose phase-shift angle is determined by a physical rotation of the HWP.
The phase-shifter corresponds to the rotation, in the plane which include the $S_3$ axis, which can be achieved by inserting 2 QWP before and after the genuine rotation in the $S_1-S_2$ plane.
In order to rotate in the $S_1-S_3$ plane, we need to apply the QWP, whose FA is aligned vertically.
This will bring the $S_3$ axis to the $S_2$ axis by the $90^{\circ}$ clock-wise rotation along the $S_1$ axis.
Then, we can apply the genuine rotator to rotate within the $S_1-S_2$ plane by using 2 HWPs.
Finally, we use another QWP, whose FA is aligned horizontally, to bring the rotated axis back to the original one by the $90^{\circ}$ anti-clock-wise rotation along the $S_1$ axis.
The amount of the rotation is determined by the rotated HWP, which is the third plate among 4 plates, such that the amount of the phase-shift angle is expected to be 4 times that of the physical rotation angle, as for a genuine rotator.

Experimental results on the inputs of $|{\rm H} \rangle$ and $|{\rm V} \rangle$ are shown in Fig. 10.
We confirm that the phase-shift vanishes without the rotation (${\it \Delta \Psi}=0$), such that the genuine phase-shifter is continuously connected to the identity operator of {\bf 1}.
This is consistent with the fact that the phase-shifter forms a sub-group in $SU(2)$.
As we rotate the HWP, the polarisation states rotated 4 times along the meridian across the Poincar\'e sphere upon the physical rotation of 1 time.

\begin{figure}[h]
\begin{center}
\includegraphics[width=8cm]{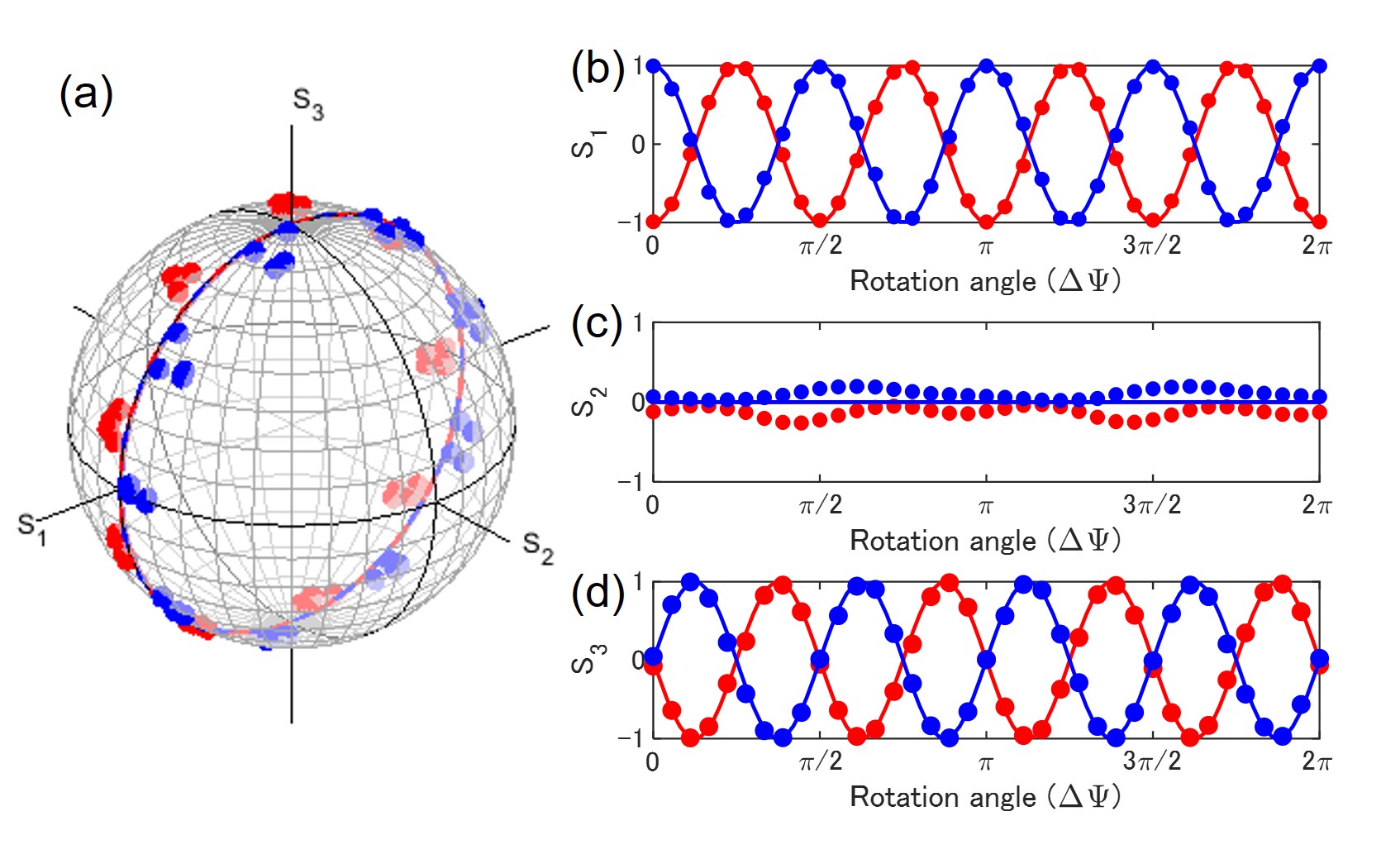}
\caption{
Phase-shifter operation by rotating a half-wave-plate for inputs of horizontally (blue) and vertically (red) polarised states.
2 quarter-wave-plates were inserted before and after the rotator operation.
(a) Trajectories of polarisation states in the Poincar\'e sphere.
(b) $S_1$, (c) $S_2$, and $S_3$ changed upon the physical rotation (${\it \Delta \Psi}$) of the half-wave-plate.
}
\end{center}
\end{figure}

\begin{figure}[h]
\begin{center}
\includegraphics[width=8cm]{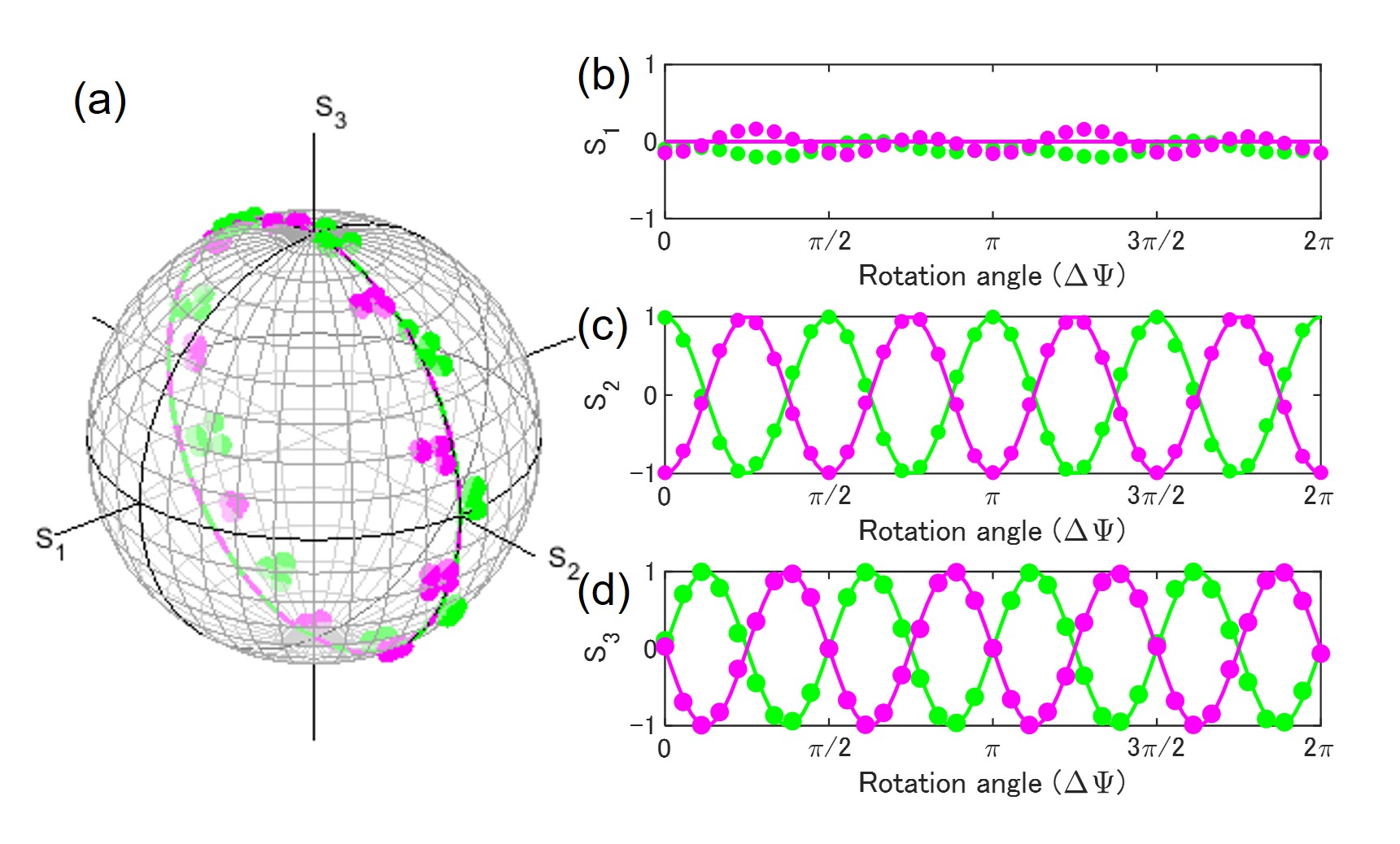}
\caption{
Phase-shifter operation by rotating a half-wave-plate for inputs of diagonally (green) and anti-diagonally (magenta) polarised states.
2 quarter-wave-plates were inserted before and after the rotator operation.
(a) Trajectories of polarisation states in the Poincar\'e sphere.
(b) $S_1$, (c) $S_2$, and $S_3$ changed upon the physical rotation (${\it \Delta \Psi}$) of the half-wave-plate.
}
\end{center}
\end{figure}

In order to rotate in the $S_2-S_3$ plane, which is more standard for a phase-shift, we need to apply the QWP, whose FA is rotated $45^{\circ}$ for the clock-wise direction.
This will bring the $S_3$ axis to the $S_1$ axis by the $90^{\circ}$ clock-wise rotation along the $S_2$ axis.
Then, we can apply the genuine rotator to rotate within the $S_1-S_2$ plane by using 2 HWPs, as before.
Finally, we use another QWP, whose FA is rotated $45^{\circ}$ for the anti-clock-wise direction to bring the rotated axis back.
This can be confirmed by calculating 
\begin{eqnarray}
&&
{\Delta}_{\rm HV}(\pi/2,\pi/2) 
{\Delta}_{\rm HV}({\it \Delta \phi},\pi) 
{\Delta}_{\rm HV}(0,\pi) 
{\Delta}_{\rm HV}(-\pi/2,\pi/2)  \nonumber \\
&&=
-\frac{1}{2}
\left (
  \begin{array}{cc}
    1 &  -i \\
    -i &   1
  \end{array}
\right)
\left (
  \begin{array}{cc}
    \cos({\it \Delta \phi}) &  -\sin({\it \Delta \phi})\\
    \sin({\it \Delta \phi}) &   \cos({\it \Delta \phi})
  \end{array}
\right)
\left (
  \begin{array}{cc}
    1 &  i \\
    i &   1
  \end{array}
\right), \nonumber \\
&&=
-
\left (
  \begin{array}{cc}
    \exp (-i{\it \Delta \phi}) & 0  \\
    0 &   \exp (+i{\it \Delta \phi})
  \end{array}
\right) \\
&&=
-
{\Delta}_{\rm HV}
(2 {\it \Delta \phi})
, 
\end{eqnarray}
which means that we can realise the proper phase-shifter, ${\Delta}_{\rm HV}(\delta)=\hat{\mathcal{D}}^{\rm HV}({\bf \hat{n}}_{1},\delta)$ with the phase-shift of $\delta=2{\it \Delta \phi}=4{\it \Delta \Psi}$, determined by physical rotation angle.

As shown in Fig. 11, we confirm the expected phase-shift for the inputs of $|{\rm D} \rangle$ and $|{\rm A} \rangle$.
Again, we confirmed that the phase-shift vanished without the rotation (${\it \Delta \Psi}=0$).
The rotation in the $S_2-S_3$ plane is quite useful especially for considering HV-bases.
By utilising this technique, one can easily realise arbitrary phase-shift in a laboratory solely by physical rotation of the wave-plates using widely available HWPs and QWPs.

\section{Discussions and conclusions}

We discuss mathematical and physical reasons why we could construct a rotator and a phase-shifter, simply from combinations of HWPs and QWPs for the perspective of Lie group.
As we have shown, the crucial point was to construct a subgroup $SO(2)$ in $SO(3)$ for spin expectation values of ${\rm S}$, represented by $\hat{\mathcal{R}}_{O_2}(2{\it \Delta \phi}) $ in Eq. (41).

This rotation keeps the $S_3$ component, such that the rotation plane is perpendicular to the $S_3$ axis.
In LR bases, this corresponds to maintain $\theta$, while changing $\phi$ to rotate along the parallel in the Poincar\'e sphere. 
In the original $SU(2)$ operator for the wavefunction, this was achieved by $\mathcal{R}_{\rm LR} (2 {\it \Delta \phi})$ of Eq. (42).
$\hat{\mathcal{R}}_{O_2}(2{\it \Delta \phi})$ and $\mathcal{R}_{\rm LR} (2 {\it \Delta \phi})$ are indeed equivalent due to the mapping of $\exp (i 2{\it \Delta \phi}) = \cos (2{\it \Delta \phi}) + i \sin (2{\it \Delta \phi})$.

Therefore, the 2-dimensional rotator is equivalent to $SO(2) \cong U(1)=\{ \exp (i {\it \phi}) | \phi \in  \mathbb{R} \}$, which forms a 1-parameter group \cite{Fulton04,Hall03,Pfeifer03,Georgi99}.
To describe the rotation along the $S_3$ axis, we do not need to use a $2 \times 2$ matrix, and 1 complex number of $\exp (i 2{\it \Delta \phi})$ is sufficient.
For fixed $\theta$ ($S_3$), the corresponding wavefunction for $U(1)$ is simply given by 
\begin{eqnarray}
 |\phi \rangle
&=&
 {\rm e}^{i \phi}
,
\end{eqnarray}
which works as a continuous basis \cite{Swanson92}, 
and the application of the U(1) rotation is given by 
\begin{eqnarray}
\mathcal{R}_{U_1} (2 {\it \Delta \phi})=\exp(i2{\it \Delta \phi}),
\end{eqnarray}
where the subscript of 3 stands for the rotation along the $S_3$ axis, such that we obtain 
\begin{eqnarray}
\mathcal{R}_{U_1} (2 {\it \Delta \phi})
 |\phi \rangle
=
 |\phi + 2{\it \Delta \phi} \rangle.
\end{eqnarray}

Consequently, we confirm that the rotator merely corresponds to the mapping of $\phi \rightarrow \phi + 2{\it \Delta \phi}$ by the $U(1)$ subgroup embedded in $SU(2)$, and the rotation along $S_3$ was achieved without affecting $\theta$.
The $U(1)$ wavefunction could be embedded to the original $SU(2)$ wavefunction in LR-bases as 
\begin{eqnarray}
 |\phi \rangle
\rightarrow
 {\rm e}^{-i\frac{\phi}{2}}  \cos (\theta/2)
 |{\rm L} \rangle
+
 {\rm e}^{+i\frac{\phi}{2}}\sin (\theta/2)
 |{\rm R} \rangle,
\end{eqnarray}
but we must be careful for using the $U(1)$ representation of $\mathcal{R}_{U_1}$ for $|\phi \rangle$  (the left-hand side of Eq. (55)) , while the $SU(2)$ representation of $\mathcal{R}_{\rm LR} (2 {\it \Delta \phi})$ must be used for $|\theta, \phi \rangle$ (the right-hand side of Eq. (55)). 
Mathematically, $SU(2)$ contains $U(1)$, such that $U(1) \subset SU(2)$ and we confirmed $O^{-}(2) \cdot \sigma_3  \cong SO(2) \cong U(1)$ to convert from the pseudo rotator to the genuine rotator.

Practically, the rotation angle in the Poincar\'e sphere is determined by the physical rotation angle, such that we can continuously change the 1-parameter in $U(1)$ by hand.
Therefore, our rotator is physical realisation of $U(1)$ for polarisation states.

Having constructed a rotator, it was straightforward to construct a phase-shifter, since we just needed to change the rotation axis by a QWP before the rotation, and bring back to the original coordinate by a $90^{\circ}$-rotated QWP from the first one after the rotation.
This corresponds to realise an $SU(2)$ rotation 
\begin{eqnarray}
\hat{\mathcal{R}}_i (\delta \phi)
&\equiv&
\hat{\mathcal{D}} ({\bf \hat{n}}_i,\delta \phi) 
=\exp 
\left (
-i 
\hat{\bm \sigma} \cdot {\bf \hat{n}}_i
\left (
\frac{\delta \phi}{2}
\right)
\right), 
\end{eqnarray}
for $i=1,2,3$, and $\hat{\mathcal{R}}_1 (\delta \phi)$ is usually called as a phase-shifter and $\hat{\mathcal{R}}_3 (\delta \phi)$ is called as a rotator.
Combining both a rotator and a phase-shifter, we can realise an arbitral rotation of the polarisation state in the Poincar\'e sphere, such that we call as a {\it Poincar\'e rotator}.
For example, we can easily construct 
\begin{eqnarray}
 |\theta,\phi \rangle
=
\hat{\mathcal{R}}_3 (\phi)
\hat{\mathcal{R}}_2 (\theta)
|{\rm L} \rangle,
\end{eqnarray}
which is suitable for LR bases.
We must be careful on the amount of expected rotation in the Poincar\'e sphere is 4 times of that of the physical rotation of HWPs.
We can also construct
\begin{eqnarray}
 |\gamma, \delta \rangle
=
\hat{\mathcal{R}}_1 (\delta)
\hat{\mathcal{R}}_3 (\gamma)
|{\rm H} \rangle,
\end{eqnarray}
which is suitable for HV-bases.

We can also realise an Euler rotation \cite{Sakurai14} 
\begin{eqnarray}
\hat{\mathcal{R}} (\alpha,\beta,\gamma)
=
\hat{\mathcal{R}}_3 (\alpha)
\hat{\mathcal{R}}_2 (\beta)
\hat{\mathcal{R}}_3 (\gamma)
\end{eqnarray}
for an arbitrary rotation in the 3-dimensional Poincar\'e sphere.
 
An advantage to use our Poincar\'e rotator is the ability that we can perform expected amount of rotation along the preferred axis without knowing the polarisation state in the input.
As we have shown theoretically and confirmed experimentally, the Poincar\'e rotator works as a subgroup of $U(1)$ upon the physical rotation, which means that the polarisation state can be controlled continuously changed from the input state.
To guarantee this, it was very important to make sure that the operation contains the identity operation of {\bf 1} to make sure that the operation is realised by a continuous change of the operation from {\bf 1}.
This is crucial requirement for a Lie group \cite{Stubhaug02,Fulton04,Hall03,Pfeifer03,Dirac30,Georgi99}, since Lie group and Lie algebra were constructed from group theoretical considerations near the operation around identities.
Consequently, by using Poincar\'e rotator, we can apply the same amount of rotation, regardless of the polarisation states of the input beam, which was not possible in a pseudo rotator configuration.
This characteristic would be useful for some applications to require a certain rotation without measuring the input state.

A Poincar\'e rotator is also useful to control the orbital angular momentum of photons \cite{Saito21}.
The left and right vortexed states are orthogonal each other, such that they form $SU(2)$ states \cite{Allen92,Padgett99,Milione11,Naidoo16,Liu17,Erhard18,Saito21,Andrews21,Angelsky21}.
A superposition states with these vortices can be controlled by a Poincar\'e rotator by adjusting the phase and amplitudes \cite{Saito21}.

So far, all theoretical considerations and experimental results are consistent with the assessment that coherent photons have an $SU(2)$ symmetry and we can apply a standard quantum mechanical prescription for an $SU(2)$ state to understand the polarisation states \cite{Stokes51,Poincare92,Jones41,Fano54,Baym69,Sakurai14,Max99,Jackson99,Yariv97,Gil16,Goldstein11,
Hecht17,Pedrotti07,Saito20a,Saito20b,Saito20c,Saito20d}.
We think that the physical origin of the macroscopic quantum coherence of polarisation is coming from the broken symmetry upon lasing threshold \cite{Saito20a,Saito20b,Saito20c,Saito20d}, such that we can treat coherent photons as a simple 2-level system to account for their spin expectation values.
The impacts of optical wave-plates could be explained by corresponding rotations in the Poincar\'e sphere \cite{Stokes51,Poincare92,Jones41,Fano54,Baym69,Sakurai14,Max99,Jackson99,Yariv97,Gil16,Goldstein11,
Hecht17,Pedrotti07,Saito20a,Saito20b,Saito20c,Saito20d}. 
We have shown that the underlying mathematical foundation for polarisation states is deeply routed in Lie group and Lie algebra.
By applying isomorphism theorems \cite{Fulton04,Hall03,Pfeifer03,Georgi99} for coherent photons, we confirmed the relationship between $SU(2)$ rotation for the wavefunction and the resultant $SO(3)$ rotation for spin expectation values.
We also found that a pseudo rotator made by a rotated half-wave-plate is describing mirror reflections and we could convert it by introducing another half-wave-plate to realise a genuine rotator by 2 plates.
This corresponds to converting $O^{-}(2)$ to $O^{+}(2) \cong SO(2)$ by $\sigma_3$.
By changing the rotation axes by quarter-wave-plates, we could also make a genuine phase-shifter, such that the arbitrary rotations can be realised by a proposed passive Poincar\'e rotator.
The implication of this work is a perspective that we can utilise the $SU(2)$ degree of freedom in coherent photons for potential quantum technologies.

\section*{Acknowledgements}
This work is supported by JSPS KAKENHI Grant Number JP 18K19958.
The author would like to express sincere thanks to Prof I. Tomita for continuous discussions and encouragements.

\bibliography{PassivePR}

\begin{thebibliography}{59}%
\makeatletter
\providecommand \@ifxundefined [1]{%
 \@ifx{#1\undefined}
}%
\providecommand \@ifnum [1]{%
 \ifnum #1\expandafter \@firstoftwo
 \else \expandafter \@secondoftwo
 \fi
}%
\providecommand \@ifx [1]{%
 \ifx #1\expandafter \@firstoftwo
 \else \expandafter \@secondoftwo
 \fi
}%
\providecommand \natexlab [1]{#1}%
\providecommand \enquote  [1]{``#1''}%
\providecommand \bibnamefont  [1]{#1}%
\providecommand \bibfnamefont [1]{#1}%
\providecommand \citenamefont [1]{#1}%
\providecommand \href@noop [0]{\@secondoftwo}%
\providecommand \href [0]{\begingroup \@sanitize@url \@href}%
\providecommand \@href[1]{\@@startlink{#1}\@@href}%
\providecommand \@@href[1]{\endgroup#1\@@endlink}%
\providecommand \@sanitize@url [0]{\catcode `\\12\catcode `\$12\catcode
  `\&12\catcode `\#12\catcode `\^12\catcode `\_12\catcode `\%12\relax}%
\providecommand \@@startlink[1]{}%
\providecommand \@@endlink[0]{}%
\providecommand \url  [0]{\begingroup\@sanitize@url \@url }%
\providecommand \@url [1]{\endgroup\@href {#1}{\urlprefix }}%
\providecommand \urlprefix  [0]{URL }%
\providecommand \Eprint [0]{\href }%
\providecommand \doibase [0]{https://doi.org/}%
\providecommand \selectlanguage [0]{\@gobble}%
\providecommand \bibinfo  [0]{\@secondoftwo}%
\providecommand \bibfield  [0]{\@secondoftwo}%
\providecommand \translation [1]{[#1]}%
\providecommand \BibitemOpen [0]{}%
\providecommand \bibitemStop [0]{}%
\providecommand \bibitemNoStop [0]{.\EOS\space}%
\providecommand \EOS [0]{\spacefactor3000\relax}%
\providecommand \BibitemShut  [1]{\csname bibitem#1\endcsname}%
\let\auto@bib@innerbib\@empty
\bibitem [{\citenamefont {Stubhaug}(2002)}]{Stubhaug02}%
  \BibitemOpen
  \bibfield  {author} {\bibinfo {author} {\bibfnamefont {A.}~\bibnamefont
  {Stubhaug}},\ }\href@noop {} {\emph {\bibinfo {title} {The Mathematician
  Sophus Lie - It was the Audacity of My Thinking}}}\ (\bibinfo  {publisher}
  {Springer-Verlag, Berlin},\ \bibinfo {year} {2002})\BibitemShut {NoStop}%
\bibitem [{\citenamefont {Fulton}\ and\ \citenamefont
  {Harris}(2004)}]{Fulton04}%
  \BibitemOpen
  \bibfield  {author} {\bibinfo {author} {\bibfnamefont {W.}~\bibnamefont
  {Fulton}}\ and\ \bibinfo {author} {\bibfnamefont {J.}~\bibnamefont
  {Harris}},\ }\href@noop {} {\emph {\bibinfo {title} {Representation Theory: A
  First Course}}}\ (\bibinfo  {publisher} {Springer, New York},\ \bibinfo
  {year} {2004})\BibitemShut {NoStop}%
\bibitem [{\citenamefont {Hall}(2003)}]{Hall03}%
  \BibitemOpen
  \bibfield  {author} {\bibinfo {author} {\bibfnamefont {B.~C.}\ \bibnamefont
  {Hall}},\ }\href@noop {} {\emph {\bibinfo {title} {Lie Groups, Lie Algebras,
  and Representations; An Elementary Introduction}}}\ (\bibinfo  {publisher}
  {Springer, Switzerland},\ \bibinfo {year} {2003})\BibitemShut {NoStop}%
\bibitem [{\citenamefont {Pfeifer}(2003)}]{Pfeifer03}%
  \BibitemOpen
  \bibfield  {author} {\bibinfo {author} {\bibfnamefont {W.}~\bibnamefont
  {Pfeifer}},\ }\href@noop {} {\emph {\bibinfo {title} {The {L}ie Algebras
  $su(N)$ An Introduction}}}\ (\bibinfo  {publisher} {Springer Basel AG,
  Berlin},\ \bibinfo {year} {2003})\BibitemShut {NoStop}%
\bibitem [{\citenamefont {Dirac}(1930)}]{Dirac30}%
  \BibitemOpen
  \bibfield  {author} {\bibinfo {author} {\bibfnamefont {P.~A.~M.}\
  \bibnamefont {Dirac}},\ }\href@noop {} {\emph {\bibinfo {title} {The
  Principle of Quantum Mechanics}}}\ (\bibinfo  {publisher} {Oxford University
  Press, Oxford},\ \bibinfo {year} {1930})\BibitemShut {NoStop}%
\bibitem [{\citenamefont {Georgi}(1999)}]{Georgi99}%
  \BibitemOpen
  \bibfield  {author} {\bibinfo {author} {\bibfnamefont {H.}~\bibnamefont
  {Georgi}},\ }\href@noop {} {\emph {\bibinfo {title} {Lie Algebras in Particle
  Physics: from Isospin to Unified Theories (Frontiers in Physics)}}}\
  (\bibinfo  {publisher} {Westview Press, Massachusetts},\ \bibinfo {year}
  {1999})\BibitemShut {NoStop}%
\bibitem [{\citenamefont {Baym}(1969)}]{Baym69}%
  \BibitemOpen
  \bibfield  {author} {\bibinfo {author} {\bibfnamefont {G.}~\bibnamefont
  {Baym}},\ }\href@noop {} {\emph {\bibinfo {title} {Lectures on Quantum
  Mechanics}}}\ (\bibinfo  {publisher} {Westview Press, New York},\ \bibinfo
  {year} {1969})\BibitemShut {NoStop}%
\bibitem [{\citenamefont {Sakurai}(1967)}]{Sakurai67}%
  \BibitemOpen
  \bibfield  {author} {\bibinfo {author} {\bibfnamefont {J.~J.}\ \bibnamefont
  {Sakurai}},\ }\href@noop {} {\emph {\bibinfo {title} {Advanced Quantum
  Mechanics}}}\ (\bibinfo  {publisher} {Addison-Wesley Publishing Company, New
  York},\ \bibinfo {year} {1967})\BibitemShut {NoStop}%
\bibitem [{\citenamefont {Sakurai}\ and\ \citenamefont
  {Napolitano}(2014)}]{Sakurai14}%
  \BibitemOpen
  \bibfield  {author} {\bibinfo {author} {\bibfnamefont {J.~J.}\ \bibnamefont
  {Sakurai}}\ and\ \bibinfo {author} {\bibfnamefont {J.~J.}\ \bibnamefont
  {Napolitano}},\ }\href@noop {} {\emph {\bibinfo {title} {Modern Quantum
  Mechanics}}}\ (\bibinfo  {publisher} {Pearson, Edinburgh},\ \bibinfo {year}
  {2014})\BibitemShut {NoStop}%
\bibitem [{\citenamefont {Nielsen}\ and\ \citenamefont
  {Chuang}(2000)}]{Nielsen00}%
  \BibitemOpen
  \bibfield  {author} {\bibinfo {author} {\bibfnamefont {M.}~\bibnamefont
  {Nielsen}}\ and\ \bibinfo {author} {\bibfnamefont {I.}~\bibnamefont
  {Chuang}},\ }\href@noop {} {\emph {\bibinfo {title} {Quantum Computation and
  Quantum Information}}}\ (\bibinfo  {publisher} {Cambridge Univ. Press,
  Cambridge},\ \bibinfo {year} {2000})\BibitemShut {NoStop}%
\bibitem [{\citenamefont {Nakamura}\ \emph {et~al.}(1999)\citenamefont
  {Nakamura}, \citenamefont {Pashkin},\ and\ \citenamefont
  {Tsai}}]{Nakamura99}%
  \BibitemOpen
  \bibfield  {author} {\bibinfo {author} {\bibfnamefont {Y.}~\bibnamefont
  {Nakamura}}, \bibinfo {author} {\bibfnamefont {Y.~A.}\ \bibnamefont
  {Pashkin}},\ and\ \bibinfo {author} {\bibfnamefont {J.~S.}\ \bibnamefont
  {Tsai}},\ }\bibfield  {title} {\bibinfo {title} {Coherent control of
  macroscopic quantum states in a single-cooper-pair box},\ }\href
  {https://doi.org/10.1038/19718} {\bibfield  {journal} {\bibinfo  {journal}
  {Nat.}\ }\textbf {\bibinfo {volume} {398}},\ \bibinfo {pages} {786} (\bibinfo
  {year} {1999})}\BibitemShut {NoStop}%
\bibitem [{\citenamefont {Koch}\ \emph {et~al.}(2007)\citenamefont {Koch},
  \citenamefont {Yu}, \citenamefont {Gambetta}, \citenamefont {Houck},
  \citenamefont {Shuster}, \citenamefont {Majer}, \citenamefont {Blais},
  \citenamefont {Devoret}, \citenamefont {Girvin},\ and\ \citenamefont
  {Schoelkopf}}]{Koch07}%
  \BibitemOpen
  \bibfield  {author} {\bibinfo {author} {\bibfnamefont {J.}~\bibnamefont
  {Koch}}, \bibinfo {author} {\bibfnamefont {T.~M.}\ \bibnamefont {Yu}},
  \bibinfo {author} {\bibfnamefont {J.}~\bibnamefont {Gambetta}}, \bibinfo
  {author} {\bibfnamefont {A.~A.}\ \bibnamefont {Houck}}, \bibinfo {author}
  {\bibfnamefont {D.~I.}\ \bibnamefont {Shuster}}, \bibinfo {author}
  {\bibfnamefont {J.}~\bibnamefont {Majer}}, \bibinfo {author} {\bibfnamefont
  {A.}~\bibnamefont {Blais}}, \bibinfo {author} {\bibfnamefont {M.~H.}\
  \bibnamefont {Devoret}}, \bibinfo {author} {\bibfnamefont {S.~M.}\
  \bibnamefont {Girvin}},\ and\ \bibinfo {author} {\bibfnamefont {R.~J.}\
  \bibnamefont {Schoelkopf}},\ }\bibfield  {title} {\bibinfo {title}
  {Charge-insensitive qubit design derived from the cooper pair box},\ }\href
  {https://doi.org/10.1103/PhysRevA.76.042319} {\bibfield  {journal} {\bibinfo
  {journal} {Phys. Rev. A}\ }\textbf {\bibinfo {volume} {76}},\ \bibinfo
  {pages} {042319} (\bibinfo {year} {2007})}\BibitemShut {NoStop}%
\bibitem [{\citenamefont {Schreier}\ \emph {et~al.}(2008)\citenamefont
  {Schreier}, \citenamefont {Houck}, \citenamefont {Koch}, \citenamefont
  {Schuster}, \citenamefont {Johnson}, \citenamefont {Chow}, \citenamefont
  {Gambetta}, \citenamefont {Majer}, \citenamefont {Frunzio}, \citenamefont
  {Devoret},\ and\ \citenamefont {Schoelkopf}}]{Schreier08}%
  \BibitemOpen
  \bibfield  {author} {\bibinfo {author} {\bibfnamefont {J.~A.}\ \bibnamefont
  {Schreier}}, \bibinfo {author} {\bibfnamefont {A.~A.}\ \bibnamefont {Houck}},
  \bibinfo {author} {\bibfnamefont {J.}~\bibnamefont {Koch}}, \bibinfo {author}
  {\bibfnamefont {D.~I.}\ \bibnamefont {Schuster}}, \bibinfo {author}
  {\bibfnamefont {B.~R.}\ \bibnamefont {Johnson}}, \bibinfo {author}
  {\bibfnamefont {J.~M.}\ \bibnamefont {Chow}}, \bibinfo {author}
  {\bibfnamefont {J.~M.}\ \bibnamefont {Gambetta}}, \bibinfo {author}
  {\bibfnamefont {J.}~\bibnamefont {Majer}}, \bibinfo {author} {\bibfnamefont
  {L.}~\bibnamefont {Frunzio}}, \bibinfo {author} {\bibfnamefont {M.~H.}\
  \bibnamefont {Devoret}},\ and\ \bibinfo {author} {\bibfnamefont {R.~J.}\
  \bibnamefont {Schoelkopf}},\ }\bibfield  {title} {\bibinfo {title}
  {Suppressing charge noise decoherence in superconducting charge qubits},\
  }\href {https://doi.org/10.1103/PhysRevB.77.180502} {\bibfield  {journal}
  {\bibinfo  {journal} {Phys. Rev. B}\ }\textbf {\bibinfo {volume} {77}},\
  \bibinfo {pages} {180502(R)} (\bibinfo {year} {2008})}\BibitemShut {NoStop}%
\bibitem [{\citenamefont {Arute}\ \emph {et~al.}(2019)\citenamefont {Arute},
  \citenamefont {Arya}, \citenamefont {Babbush}, \citenamefont {Bacon},
  \citenamefont {Bardin},\ and\ \citenamefont {{\it et. al.}}}]{Arute19}%
  \BibitemOpen
  \bibfield  {author} {\bibinfo {author} {\bibfnamefont {F.}~\bibnamefont
  {Arute}}, \bibinfo {author} {\bibfnamefont {K.}~\bibnamefont {Arya}},
  \bibinfo {author} {\bibfnamefont {R.}~\bibnamefont {Babbush}}, \bibinfo
  {author} {\bibfnamefont {D.}~\bibnamefont {Bacon}}, \bibinfo {author}
  {\bibfnamefont {J.~C.}\ \bibnamefont {Bardin}},\ and\ \bibinfo {author}
  {\bibnamefont {{\it et. al.}}},\ }\bibfield  {title} {\bibinfo {title}
  {Quantum supremacy using a programmable superconducting processor},\ }\href
  {https://doi.org/10.1038/s41586-019-1666-5} {\bibfield  {journal} {\bibinfo
  {journal} {Nat.}\ }\textbf {\bibinfo {volume} {574}},\ \bibinfo {pages} {505}
  (\bibinfo {year} {2019})}\BibitemShut {NoStop}%
\bibitem [{\citenamefont {Bruzewicz}\ \emph {et~al.}(2019)\citenamefont
  {Bruzewicz}, \citenamefont {Chiaverini}, \citenamefont {McConnell},\ and\
  \citenamefont {Saga}}]{Bruzewicz19}%
  \BibitemOpen
  \bibfield  {author} {\bibinfo {author} {\bibfnamefont {C.~D.}\ \bibnamefont
  {Bruzewicz}}, \bibinfo {author} {\bibfnamefont {J.}~\bibnamefont
  {Chiaverini}}, \bibinfo {author} {\bibfnamefont {R.}~\bibnamefont
  {McConnell}},\ and\ \bibinfo {author} {\bibfnamefont {J.~M.}\ \bibnamefont
  {Saga}},\ }\bibfield  {title} {\bibinfo {title} {Trapped-ion quantum
  computing: Progress and challenges},\ }\href
  {https://doi.org/10.1063/1.5088164} {\bibfield  {journal} {\bibinfo
  {journal} {Appl. Phys. Rev.}\ }\textbf {\bibinfo {volume} {6}},\ \bibinfo
  {pages} {021314} (\bibinfo {year} {2019})}\BibitemShut {NoStop}%
\bibitem [{\citenamefont {Pino}\ \emph {et~al.}(2021)\citenamefont {Pino},
  \citenamefont {Dreiling}, \citenamefont {Figgatt}, \citenamefont {Gaebler},
  \citenamefont {Moses}, \citenamefont {Allman}, \citenamefont {Baldwin},
  \citenamefont {Foss-Feig}, \citenamefont {Hayes}, \citenamefont {Mayer},
  \citenamefont {Ryan-Anderson},\ and\ \citenamefont {Neyenhuis}}]{Pino21}%
  \BibitemOpen
  \bibfield  {author} {\bibinfo {author} {\bibfnamefont {J.~M.}\ \bibnamefont
  {Pino}}, \bibinfo {author} {\bibfnamefont {J.~M.}\ \bibnamefont {Dreiling}},
  \bibinfo {author} {\bibfnamefont {C.}~\bibnamefont {Figgatt}}, \bibinfo
  {author} {\bibfnamefont {J.~P.}\ \bibnamefont {Gaebler}}, \bibinfo {author}
  {\bibfnamefont {S.~A.}\ \bibnamefont {Moses}}, \bibinfo {author}
  {\bibfnamefont {M.~S.}\ \bibnamefont {Allman}}, \bibinfo {author}
  {\bibfnamefont {C.~H.}\ \bibnamefont {Baldwin}}, \bibinfo {author}
  {\bibfnamefont {M.}~\bibnamefont {Foss-Feig}}, \bibinfo {author}
  {\bibfnamefont {D.}~\bibnamefont {Hayes}}, \bibinfo {author} {\bibfnamefont
  {K.}~\bibnamefont {Mayer}}, \bibinfo {author} {\bibfnamefont
  {C.}~\bibnamefont {Ryan-Anderson}},\ and\ \bibinfo {author} {\bibfnamefont
  {B.}~\bibnamefont {Neyenhuis}},\ }\bibfield  {title} {\bibinfo {title}
  {Demonstration of the trapped-ion quantum {CCD} computer architecture},\
  }\href {https://doi.org/10.1038/s41586-021-03318-4} {\bibfield  {journal}
  {\bibinfo  {journal} {Nat.}\ }\textbf {\bibinfo {volume} {592}},\ \bibinfo
  {pages} {209} (\bibinfo {year} {2021})}\BibitemShut {NoStop}%
\bibitem [{\citenamefont {O'Brien}\ \emph {et~al.}(2003)\citenamefont
  {O'Brien}, \citenamefont {Pryde},\ and\ \citenamefont {White}}]{OBrien03}%
  \BibitemOpen
  \bibfield  {author} {\bibinfo {author} {\bibfnamefont {J.}~\bibnamefont
  {O'Brien}}, \bibinfo {author} {\bibfnamefont {G.}~\bibnamefont {Pryde}},\
  and\ \bibinfo {author} {\bibfnamefont {A.}~\bibnamefont {White}},\ }\bibfield
   {title} {\bibinfo {title} {Demonstration of an all-optical quantum
  controlled-{NOT} gate},\ }\href {https://doi.org/10.1038/nature02054}
  {\bibfield  {journal} {\bibinfo  {journal} {Nat.}\ }\textbf {\bibinfo
  {volume} {426}},\ \bibinfo {pages} {264} (\bibinfo {year}
  {2003})}\BibitemShut {NoStop}%
\bibitem [{\citenamefont {Peruzzo}\ \emph {et~al.}(2014)\citenamefont
  {Peruzzo}, \citenamefont {McClean}, \citenamefont {Shadbolt}, \citenamefont
  {Yung}, \citenamefont {Zhou}, \citenamefont {Love}, \citenamefont
  {Aspuru-Guzik},\ and\ \citenamefont {O'Brian}}]{Peruzzo14}%
  \BibitemOpen
  \bibfield  {author} {\bibinfo {author} {\bibfnamefont {A.}~\bibnamefont
  {Peruzzo}}, \bibinfo {author} {\bibfnamefont {J.}~\bibnamefont {McClean}},
  \bibinfo {author} {\bibfnamefont {P.}~\bibnamefont {Shadbolt}}, \bibinfo
  {author} {\bibfnamefont {M.~H.}\ \bibnamefont {Yung}}, \bibinfo {author}
  {\bibfnamefont {X.~Q.}\ \bibnamefont {Zhou}}, \bibinfo {author}
  {\bibfnamefont {P.~J.}\ \bibnamefont {Love}}, \bibinfo {author}
  {\bibfnamefont {A.}~\bibnamefont {Aspuru-Guzik}},\ and\ \bibinfo {author}
  {\bibfnamefont {J.~L.}\ \bibnamefont {O'Brian}},\ }\bibfield  {title}
  {\bibinfo {title} {A variational eigenvalue solver on a photonic quantum
  processor},\ }\href {https://doi.org/10.1038/ncomms5213} {\bibfield
  {journal} {\bibinfo  {journal} {Nat. Commun.}\ }\textbf {\bibinfo {volume}
  {5}},\ \bibinfo {pages} {4213} (\bibinfo {year} {2014})}\BibitemShut
  {NoStop}%
\bibitem [{\citenamefont {Silverstone}\ \emph {et~al.}(2016)\citenamefont
  {Silverstone}, \citenamefont {Bonneau}, \citenamefont {O'Brien},\ and\
  \citenamefont {Thompson}}]{Silverstone16}%
  \BibitemOpen
  \bibfield  {author} {\bibinfo {author} {\bibfnamefont {J.~W.}\ \bibnamefont
  {Silverstone}}, \bibinfo {author} {\bibfnamefont {D.}~\bibnamefont
  {Bonneau}}, \bibinfo {author} {\bibfnamefont {J.~L.}\ \bibnamefont
  {O'Brien}},\ and\ \bibinfo {author} {\bibfnamefont {M.~G.}\ \bibnamefont
  {Thompson}},\ }\bibfield  {title} {\bibinfo {title} {Silicon quantum
  photonics},\ }\href {https://doi.org/10.1109/JSTQE.2016.2573218} {\bibfield
  {journal} {\bibinfo  {journal} {IEEE J. Sel. Top. Quantum Electron.}\
  }\textbf {\bibinfo {volume} {22}},\ \bibinfo {pages} {390} (\bibinfo {year}
  {2016})}\BibitemShut {NoStop}%
\bibitem [{\citenamefont {Takeda}\ and\ \citenamefont
  {Furusawa}(2017)}]{Takeda17}%
  \BibitemOpen
  \bibfield  {author} {\bibinfo {author} {\bibfnamefont {S.}~\bibnamefont
  {Takeda}}\ and\ \bibinfo {author} {\bibfnamefont {A.}~\bibnamefont
  {Furusawa}},\ }\bibfield  {title} {\bibinfo {title} {Universal quantum
  computing with measurement-induced continuous-variable gate sequence in a
  loop-based architecture},\ }\href
  {https://doi.org/10.1103/PhysRevLett.119.120504} {\bibfield  {journal}
  {\bibinfo  {journal} {Phys. Rev. Lett.}\ }\textbf {\bibinfo {volume} {119}},\
  \bibinfo {pages} {120504} (\bibinfo {year} {2017})}\BibitemShut {NoStop}%
\bibitem [{\citenamefont {Lee}\ \emph {et~al.}(2020)\citenamefont {Lee},
  \citenamefont {Tsuchiya}, \citenamefont {Shinkai}, \citenamefont {Kanno},
  \citenamefont {Mine}, \citenamefont {Takahama}, \citenamefont {Mizokuchi},
  \citenamefont {Kodera}, \citenamefont {Hisamoto},\ and\ \citenamefont
  {Mizuno}}]{Lee20}%
  \BibitemOpen
  \bibfield  {author} {\bibinfo {author} {\bibfnamefont {N.}~\bibnamefont
  {Lee}}, \bibinfo {author} {\bibfnamefont {R.}~\bibnamefont {Tsuchiya}},
  \bibinfo {author} {\bibfnamefont {G.}~\bibnamefont {Shinkai}}, \bibinfo
  {author} {\bibfnamefont {Y.}~\bibnamefont {Kanno}}, \bibinfo {author}
  {\bibfnamefont {T.}~\bibnamefont {Mine}}, \bibinfo {author} {\bibfnamefont
  {T.}~\bibnamefont {Takahama}}, \bibinfo {author} {\bibfnamefont
  {R.}~\bibnamefont {Mizokuchi}}, \bibinfo {author} {\bibfnamefont
  {T.}~\bibnamefont {Kodera}}, \bibinfo {author} {\bibfnamefont
  {D.}~\bibnamefont {Hisamoto}},\ and\ \bibinfo {author} {\bibfnamefont
  {H.}~\bibnamefont {Mizuno}},\ }\bibfield  {title} {\bibinfo {title}
  {Enhancing electrostatic coupling in silicon quantum dot array by dual gate
  oxide thickness for large-scale integration},\ }\href
  {https://doi.org/10.1063/1.5141522} {\bibfield  {journal} {\bibinfo
  {journal} {Appl. Phys. Lett.}\ }\textbf {\bibinfo {volume} {116}},\ \bibinfo
  {pages} {162106} (\bibinfo {year} {2020})}\BibitemShut {NoStop}%
\bibitem [{\citenamefont {Xue}\ \emph {et~al.}(2021)\citenamefont {Xue},
  \citenamefont {Patra}, \citenamefont {v.~Dijk}, \citenamefont {Samkharadze},
  \citenamefont {Subramanian}, \citenamefont {Corna}, \citenamefont {Wuetz},
  \citenamefont {Jeon}, \citenamefont {Sheikh}, \citenamefont
  {Juarez-Hernandez}, \citenamefont {Esparza}, \citenamefont {Rampurawala},
  \citenamefont {Carlton}, \citenamefont {Ravikumar}, \citenamefont {Nieva},
  \citenamefont {Kim}, \citenamefont {Lee}, \citenamefont {Sammak},
  \citenamefont {Scappucci}, \citenamefont {Veldhorst}, \citenamefont
  {Sebastiano}, \citenamefont {Babaie}, \citenamefont {Pellerano},
  \citenamefont {Charbon},\ and\ \citenamefont {Vandersypen}}]{Xue21}%
  \BibitemOpen
  \bibfield  {author} {\bibinfo {author} {\bibfnamefont {X.}~\bibnamefont
  {Xue}}, \bibinfo {author} {\bibfnamefont {B.}~\bibnamefont {Patra}}, \bibinfo
  {author} {\bibfnamefont {J.~P.~G.}\ \bibnamefont {v.~Dijk}}, \bibinfo
  {author} {\bibfnamefont {N.}~\bibnamefont {Samkharadze}}, \bibinfo {author}
  {\bibfnamefont {S.}~\bibnamefont {Subramanian}}, \bibinfo {author}
  {\bibfnamefont {A.}~\bibnamefont {Corna}}, \bibinfo {author} {\bibfnamefont
  {B.~P.}\ \bibnamefont {Wuetz}}, \bibinfo {author} {\bibfnamefont
  {C.}~\bibnamefont {Jeon}}, \bibinfo {author} {\bibfnamefont {F.}~\bibnamefont
  {Sheikh}}, \bibinfo {author} {\bibfnamefont {E.}~\bibnamefont
  {Juarez-Hernandez}}, \bibinfo {author} {\bibfnamefont {B.~P.}\ \bibnamefont
  {Esparza}}, \bibinfo {author} {\bibfnamefont {H.}~\bibnamefont
  {Rampurawala}}, \bibinfo {author} {\bibfnamefont {B.}~\bibnamefont
  {Carlton}}, \bibinfo {author} {\bibfnamefont {S.}~\bibnamefont {Ravikumar}},
  \bibinfo {author} {\bibfnamefont {C.}~\bibnamefont {Nieva}}, \bibinfo
  {author} {\bibfnamefont {S.}~\bibnamefont {Kim}}, \bibinfo {author}
  {\bibfnamefont {H.~J.}\ \bibnamefont {Lee}}, \bibinfo {author} {\bibfnamefont
  {A.}~\bibnamefont {Sammak}}, \bibinfo {author} {\bibfnamefont
  {G.}~\bibnamefont {Scappucci}}, \bibinfo {author} {\bibfnamefont
  {M.}~\bibnamefont {Veldhorst}}, \bibinfo {author} {\bibfnamefont
  {F.}~\bibnamefont {Sebastiano}}, \bibinfo {author} {\bibfnamefont
  {M.}~\bibnamefont {Babaie}}, \bibinfo {author} {\bibfnamefont
  {S.}~\bibnamefont {Pellerano}}, \bibinfo {author} {\bibfnamefont
  {E.}~\bibnamefont {Charbon}},\ and\ \bibinfo {author} {\bibfnamefont
  {L.~M.~K.}\ \bibnamefont {Vandersypen}},\ }\bibfield  {title} {\bibinfo
  {title} {{CMOS}-based cryogenic control of silicon quantum circuits},\ }\href
  {https://doi.org/10.1038/s41586-021-03469-4} {\bibfield  {journal} {\bibinfo
  {journal} {Nat.}\ }\textbf {\bibinfo {volume} {593}},\ \bibinfo {pages} {205}
  (\bibinfo {year} {2021})}\BibitemShut {NoStop}%
\bibitem [{\citenamefont {Preskill}(2018)}]{Preskill18}%
  \BibitemOpen
  \bibfield  {author} {\bibinfo {author} {\bibfnamefont {J.}~\bibnamefont
  {Preskill}},\ }\bibfield  {title} {\bibinfo {title} {Quantum computing in the
  nisq era and beyond},\ }\href {https://doi.org/10.22331/q-2018-08-06-79}
  {\bibfield  {journal} {\bibinfo  {journal} {Quantum}\ }\textbf {\bibinfo
  {volume} {2}},\ \bibinfo {pages} {79} (\bibinfo {year} {2018})}\BibitemShut
  {NoStop}%
\bibitem [{\citenamefont {Caldeira}\ and\ \citenamefont
  {Leggett}(1981)}]{Caldeira81}%
  \BibitemOpen
  \bibfield  {author} {\bibinfo {author} {\bibfnamefont {A.~O.}\ \bibnamefont
  {Caldeira}}\ and\ \bibinfo {author} {\bibfnamefont {A.~J.}\ \bibnamefont
  {Leggett}},\ }\bibfield  {title} {\bibinfo {title} {Influence of dissipation
  on quantum tunneling in macroscopic systems},\ }\href
  {https://doi.org/10.1103/PhysRevLett.46.211} {\bibfield  {journal} {\bibinfo
  {journal} {Phys. Rev. Lett.}\ }\textbf {\bibinfo {volume} {46}},\ \bibinfo
  {pages} {211} (\bibinfo {year} {1981})}\BibitemShut {NoStop}%
\bibitem [{\citenamefont {Born}\ and\ \citenamefont {Wolf}(1999)}]{Max99}%
  \BibitemOpen
  \bibfield  {author} {\bibinfo {author} {\bibfnamefont {M.}~\bibnamefont
  {Born}}\ and\ \bibinfo {author} {\bibfnamefont {E.}~\bibnamefont {Wolf}},\
  }\href@noop {} {\emph {\bibinfo {title} {Principles of Optics}}}\ (\bibinfo
  {publisher} {Cambridge University Press, Cambridge},\ \bibinfo {year}
  {1999})\BibitemShut {NoStop}%
\bibitem [{\citenamefont {Jackson}(1999)}]{Jackson99}%
  \BibitemOpen
  \bibfield  {author} {\bibinfo {author} {\bibfnamefont {J.~D.}\ \bibnamefont
  {Jackson}},\ }\href@noop {} {\emph {\bibinfo {title} {Classical
  Electrodynamics}}}\ (\bibinfo  {publisher} {John Wiley \& Sons, New York},\
  \bibinfo {year} {1999})\BibitemShut {NoStop}%
\bibitem [{\citenamefont {Yariv}\ and\ \citenamefont {Yeh}(1997)}]{Yariv97}%
  \BibitemOpen
  \bibfield  {author} {\bibinfo {author} {\bibfnamefont {Y.}~\bibnamefont
  {Yariv}}\ and\ \bibinfo {author} {\bibfnamefont {P.}~\bibnamefont {Yeh}},\
  }\href@noop {} {\emph {\bibinfo {title} {Photonics: optical electronics in
  modern communications}}}\ (\bibinfo  {publisher} {Oxford University Press,
  Oxford},\ \bibinfo {year} {1997})\BibitemShut {NoStop}%
\bibitem [{\citenamefont {Gil}\ and\ \citenamefont {Ossikovski}(2016)}]{Gil16}%
  \BibitemOpen
  \bibfield  {author} {\bibinfo {author} {\bibfnamefont {J.~J.}\ \bibnamefont
  {Gil}}\ and\ \bibinfo {author} {\bibfnamefont {R.}~\bibnamefont
  {Ossikovski}},\ }\href@noop {} {\emph {\bibinfo {title} {Polarized Light and
  the Mueller Matrix Approach}}}\ (\bibinfo  {publisher} {CRC Press, London},\
  \bibinfo {year} {2016})\BibitemShut {NoStop}%
\bibitem [{\citenamefont {Goldstein}(2011)}]{Goldstein11}%
  \BibitemOpen
  \bibfield  {author} {\bibinfo {author} {\bibfnamefont {D.~H.}\ \bibnamefont
  {Goldstein}},\ }\href@noop {} {\emph {\bibinfo {title} {Polarized Light}}}\
  (\bibinfo  {publisher} {CRC Press, London},\ \bibinfo {year}
  {2011})\BibitemShut {NoStop}%
\bibitem [{\citenamefont {Jones}(1941)}]{Jones41}%
  \BibitemOpen
  \bibfield  {author} {\bibinfo {author} {\bibfnamefont {R.~C.}\ \bibnamefont
  {Jones}},\ }\bibfield  {title} {\bibinfo {title} {A new calculus for the
  treatment of optical systems i. description and discussion of the calculus},\
  }\href {https://doi.org/10.1364/JOSA.31.000488} {\bibfield  {journal}
  {\bibinfo  {journal} {J. Opt. Soc. Am.}\ }\textbf {\bibinfo {volume} {31}},\
  \bibinfo {pages} {488} (\bibinfo {year} {1941})}\BibitemShut {NoStop}%
\bibitem [{\citenamefont {Fano}(1954)}]{Fano54}%
  \BibitemOpen
  \bibfield  {author} {\bibinfo {author} {\bibfnamefont {U.}~\bibnamefont
  {Fano}},\ }\bibfield  {title} {\bibinfo {title} {A stokes-parameter technique
  for the treatment of polarization in quantum mechanics},\ }\href@noop {}
  {\bibfield  {journal} {\bibinfo  {journal} {Phy. Rev.}\ }\textbf {\bibinfo
  {volume} {93}},\ \bibinfo {pages} {121} (\bibinfo {year} {1954})}\BibitemShut
  {NoStop}%
\bibitem [{\citenamefont {Stokes}(1851)}]{Stokes51}%
  \BibitemOpen
  \bibfield  {author} {\bibinfo {author} {\bibfnamefont {G.~G.}\ \bibnamefont
  {Stokes}},\ }\bibfield  {title} {\bibinfo {title} {On the composition and
  resolution of streams of polarized light from different sources},\
  }\href@noop {} {\bibfield  {journal} {\bibinfo  {journal} {Trans. Cambridge
  Phil. Soc.}\ }\textbf {\bibinfo {volume} {9}},\ \bibinfo {pages} {399}
  (\bibinfo {year} {1851})}\BibitemShut {NoStop}%
\bibitem [{\citenamefont {Poincar$\rm\acute{e}$}(1892)}]{Poincare92}%
  \BibitemOpen
  \bibfield  {author} {\bibinfo {author} {\bibfnamefont {J.~H.}\ \bibnamefont
  {Poincar$\rm\acute{e}$}},\ }\href@noop {} {\emph {\bibinfo {title}
  {Th$\rm\acute{e}$orie math$\rm\acute{e}$matique de la
  lumi$\rm\grave{e}$re}}}\ (\bibinfo  {publisher} {G. Carr$\rm\acute{e}$},\
  \bibinfo {year} {1892})\BibitemShut {NoStop}%
\bibitem [{\citenamefont {Plank}(1900)}]{Plank00}%
  \BibitemOpen
  \bibfield  {author} {\bibinfo {author} {\bibfnamefont {M.}~\bibnamefont
  {Plank}},\ }\bibfield  {title} {\bibinfo {title} {On the theory of the energy
  distribution law of the normal spectrum},\ }\href@noop {} {\bibfield
  {journal} {\bibinfo  {journal} {Verhandl. Dtsch. Phys. Ges.}\ }\textbf
  {\bibinfo {volume} {2}},\ \bibinfo {pages} {237} (\bibinfo {year}
  {1900})}\BibitemShut {NoStop}%
\bibitem [{\citenamefont {Einstein}(1905)}]{Einstein05}%
  \BibitemOpen
  \bibfield  {author} {\bibinfo {author} {\bibfnamefont {A.}~\bibnamefont
  {Einstein}},\ }\bibfield  {title} {\bibinfo {title} {Concerning an heuristic
  point of view toward the emission and transformation of light},\ }\href@noop
  {} {\bibfield  {journal} {\bibinfo  {journal} {Ann. Phys.}\ }\textbf
  {\bibinfo {volume} {17}},\ \bibinfo {pages} {132} (\bibinfo {year}
  {1905})}\BibitemShut {NoStop}%
\bibitem [{\citenamefont {Bohr}(1913)}]{Bohr13}%
  \BibitemOpen
  \bibfield  {author} {\bibinfo {author} {\bibfnamefont {N.}~\bibnamefont
  {Bohr}},\ }\bibfield  {title} {\bibinfo {title} {The spectra of helium and
  hydrogen},\ }\href {https://doi.org/10.1038/092231d0} {\bibfield  {journal}
  {\bibinfo  {journal} {Nature}\ }\textbf {\bibinfo {volume} {92}},\ \bibinfo
  {pages} {231} (\bibinfo {year} {1913})}\BibitemShut {NoStop}%
\bibitem [{\citenamefont {Dirac}(1928)}]{Dirac28}%
  \BibitemOpen
  \bibfield  {author} {\bibinfo {author} {\bibfnamefont {P.~A.~M.}\
  \bibnamefont {Dirac}},\ }\bibfield  {title} {\bibinfo {title} {The quantum
  theory of the electron},\ }\href@noop {} {\bibfield  {journal} {\bibinfo
  {journal} {Proc. R. Sco. Lond. A}\ }\textbf {\bibinfo {volume} {1117}},\
  \bibinfo {pages} {610} (\bibinfo {year} {1928})}\BibitemShut {NoStop}%
\bibitem [{\citenamefont {Abrikosov}\ \emph {et~al.}(1975)\citenamefont
  {Abrikosov}, \citenamefont {Gorkov},\ and\ \citenamefont
  {Dzyaloshinski}}]{Abrikosov75}%
  \BibitemOpen
  \bibfield  {author} {\bibinfo {author} {\bibfnamefont {A.~A.}\ \bibnamefont
  {Abrikosov}}, \bibinfo {author} {\bibfnamefont {L.~P.}\ \bibnamefont
  {Gorkov}},\ and\ \bibinfo {author} {\bibfnamefont {I.~E.}\ \bibnamefont
  {Dzyaloshinski}},\ }\href@noop {} {\emph {\bibinfo {title} {Methods of
  Quantum Field Thoery in Statistical Physics}}}\ (\bibinfo  {publisher}
  {Dover, New York},\ \bibinfo {year} {1975})\BibitemShut {NoStop}%
\bibitem [{\citenamefont {Fetter}\ and\ \citenamefont
  {Walecka}(2003)}]{Fetter03}%
  \BibitemOpen
  \bibfield  {author} {\bibinfo {author} {\bibfnamefont {A.~L.}\ \bibnamefont
  {Fetter}}\ and\ \bibinfo {author} {\bibfnamefont {J.~D.}\ \bibnamefont
  {Walecka}},\ }\href@noop {} {\emph {\bibinfo {title} {Quantum Theory of
  Many-Particle Systems}}}\ (\bibinfo  {publisher} {Dover, New York},\ \bibinfo
  {year} {2003})\BibitemShut {NoStop}%
\bibitem [{\citenamefont {Weinberg}(2005)}]{Weinberg05}%
  \BibitemOpen
  \bibfield  {author} {\bibinfo {author} {\bibfnamefont {S.}~\bibnamefont
  {Weinberg}},\ }\href@noop {} {\emph {\bibinfo {title} {The Quantum Theory of
  Fields: Foundations volume 1}}}\ (\bibinfo  {publisher} {Cambridge University
  Press, Cambridge},\ \bibinfo {year} {2005})\BibitemShut {NoStop}%
\bibitem [{\citenamefont {Fox}(2006)}]{Fox06}%
  \BibitemOpen
  \bibfield  {author} {\bibinfo {author} {\bibfnamefont {M.}~\bibnamefont
  {Fox}},\ }\href@noop {} {\emph {\bibinfo {title} {Quantum Optics: An
  Introduction}}}\ (\bibinfo  {publisher} {Oxford University Press, Oxford},\
  \bibinfo {year} {2006})\BibitemShut {NoStop}%
\bibitem [{\citenamefont {Parker}(2005)}]{Parker05}%
  \BibitemOpen
  \bibfield  {author} {\bibinfo {author} {\bibfnamefont {M.~A.}\ \bibnamefont
  {Parker}},\ }\href@noop {} {\emph {\bibinfo {title} {Physics of
  Optoelectronics}}}\ (\bibinfo  {publisher} {Tylor \& Francis},\ \bibinfo
  {year} {2005})\BibitemShut {NoStop}%
\bibitem [{\citenamefont {Altland}\ and\ \citenamefont
  {Simons}(2010)}]{Altland10}%
  \BibitemOpen
  \bibfield  {author} {\bibinfo {author} {\bibfnamefont {A.}~\bibnamefont
  {Altland}}\ and\ \bibinfo {author} {\bibfnamefont {B.}~\bibnamefont
  {Simons}},\ }\href@noop {} {\emph {\bibinfo {title} {Condensed Matter Field
  Theory}}}\ (\bibinfo  {publisher} {Cambridge University Press, Cambridge},\
  \bibinfo {year} {2010})\BibitemShut {NoStop}%
\bibitem [{\citenamefont {Hecht}(2017)}]{Hecht17}%
  \BibitemOpen
  \bibfield  {author} {\bibinfo {author} {\bibfnamefont {E.}~\bibnamefont
  {Hecht}},\ }\href@noop {} {\emph {\bibinfo {title} {Optics}}}\ (\bibinfo
  {publisher} {Pearson Education, Essex},\ \bibinfo {year} {2017})\BibitemShut
  {NoStop}%
\bibitem [{\citenamefont {Pedrotti}\ \emph {et~al.}(2007)\citenamefont
  {Pedrotti}, \citenamefont {Pedrotti},\ and\ \citenamefont
  {Pedrotti}}]{Pedrotti07}%
  \BibitemOpen
  \bibfield  {author} {\bibinfo {author} {\bibfnamefont {F.~L.}\ \bibnamefont
  {Pedrotti}}, \bibinfo {author} {\bibfnamefont {L.~M.}\ \bibnamefont
  {Pedrotti}},\ and\ \bibinfo {author} {\bibfnamefont {L.~S.}\ \bibnamefont
  {Pedrotti}},\ }\href@noop {} {\emph {\bibinfo {title} {Introduction to
  Optics}}}\ (\bibinfo  {publisher} {Pearson Education, New York},\ \bibinfo
  {year} {2007})\BibitemShut {NoStop}%
\bibitem [{\citenamefont {Saito}(shed{\natexlab{a}})}]{Saito20a}%
  \BibitemOpen
  \bibfield  {author} {\bibinfo {author} {\bibfnamefont {S.}~\bibnamefont
  {Saito}},\ }\bibfield  {title} {\bibinfo {title} {Spin of photons: Nature of
  polarisation},\ }\href@noop {} {\  (\bibinfo {year}
  {unpublished}{\natexlab{a}})}\BibitemShut {NoStop}%
\bibitem [{\citenamefont {Saito}(shed{\natexlab{b}})}]{Saito20b}%
  \BibitemOpen
  \bibfield  {author} {\bibinfo {author} {\bibfnamefont {S.}~\bibnamefont
  {Saito}},\ }\bibfield  {title} {\bibinfo {title} {Quantum commutation
  relationship for photonic orbital angular momentum},\ }\href@noop {} {\
  (\bibinfo {year} {unpublished}{\natexlab{b}})}\BibitemShut {NoStop}%
\bibitem [{\citenamefont {Saito}(shed{\natexlab{c}})}]{Saito20c}%
  \BibitemOpen
  \bibfield  {author} {\bibinfo {author} {\bibfnamefont {S.}~\bibnamefont
  {Saito}},\ }\bibfield  {title} {\bibinfo {title} {Spin and orbital angular
  momentum of coherent photons in a waveguide},\ }\href@noop {} {\  (\bibinfo
  {year} {unpublished}{\natexlab{c}})}\BibitemShut {NoStop}%
\bibitem [{\citenamefont {Saito}(shed{\natexlab{d}})}]{Saito20d}%
  \BibitemOpen
  \bibfield  {author} {\bibinfo {author} {\bibfnamefont {S.}~\bibnamefont
  {Saito}},\ }\bibfield  {title} {\bibinfo {title} {Dirac equation for photons:
  Origin of polarisation},\ }\href@noop {} {\  (\bibinfo {year}
  {unpublished}{\natexlab{d}})}\BibitemShut {NoStop}%
\bibitem [{\citenamefont {Swanson}(1992)}]{Swanson92}%
  \BibitemOpen
  \bibfield  {author} {\bibinfo {author} {\bibfnamefont {M.~S.}\ \bibnamefont
  {Swanson}},\ }\href@noop {} {\emph {\bibinfo {title} {Path Integrals and
  Quantum Processes}}}\ (\bibinfo  {publisher} {Academic press, London},\
  \bibinfo {year} {1992})\BibitemShut {NoStop}%
\bibitem [{\citenamefont {Saito}(2021)}]{Saito21}%
  \BibitemOpen
  \bibfield  {author} {\bibinfo {author} {\bibfnamefont {S.}~\bibnamefont
  {Saito}},\ }\bibfield  {title} {\bibinfo {title} {Poincar\'e rotator for
  vortexed photons},\ }\href {https://doi.org/10.3389/fphy.2021.646228}
  {\bibfield  {journal} {\bibinfo  {journal} {Front. Phys.}\ }\textbf {\bibinfo
  {volume} {9}},\ \bibinfo {pages} {646228} (\bibinfo {year}
  {2021})}\BibitemShut {NoStop}%
\bibitem [{\citenamefont {Allen}\ \emph {et~al.}(1992)\citenamefont {Allen},
  \citenamefont {Beijersbergen}, \citenamefont {Spreeuw},\ and\ \citenamefont
  {Woerdman}}]{Allen92}%
  \BibitemOpen
  \bibfield  {author} {\bibinfo {author} {\bibfnamefont {L.}~\bibnamefont
  {Allen}}, \bibinfo {author} {\bibfnamefont {M.~W.}\ \bibnamefont
  {Beijersbergen}}, \bibinfo {author} {\bibfnamefont {R.~J.~C.}\ \bibnamefont
  {Spreeuw}},\ and\ \bibinfo {author} {\bibfnamefont {J.~P.}\ \bibnamefont
  {Woerdman}},\ }\bibfield  {title} {\bibinfo {title} {Orbital angular momentum
  of light and the transformation of {L}aguerre-{G}aussian laser modes},\
  }\href {https://doi.org/10.1103/PhysRevA.45.8185} {\bibfield  {journal}
  {\bibinfo  {journal} {Phys. Rev. A}\ }\textbf {\bibinfo {volume} {45}},\
  \bibinfo {pages} {8185} (\bibinfo {year} {1992})}\BibitemShut {NoStop}%
\bibitem [{\citenamefont {Padgett}\ and\ \citenamefont
  {Courtial}(1999)}]{Padgett99}%
  \BibitemOpen
  \bibfield  {author} {\bibinfo {author} {\bibfnamefont {M.~J.}\ \bibnamefont
  {Padgett}}\ and\ \bibinfo {author} {\bibfnamefont {J.}~\bibnamefont
  {Courtial}},\ }\bibfield  {title} {\bibinfo {title}
  {Poincar$\rm\acute{e}$-sphere equivalent for light beams containing orbital
  angular momentum},\ }\href {https://doi.org/10.1364/OL.24.000430} {\bibfield
  {journal} {\bibinfo  {journal} {Opt. Lett.}\ }\textbf {\bibinfo {volume}
  {24}},\ \bibinfo {pages} {430} (\bibinfo {year} {1999})}\BibitemShut
  {NoStop}%
\bibitem [{\citenamefont {Milione}\ \emph {et~al.}(2011)\citenamefont
  {Milione}, \citenamefont {Sztul}, \citenamefont {Nolan},\ and\ \citenamefont
  {Alfano}}]{Milione11}%
  \BibitemOpen
  \bibfield  {author} {\bibinfo {author} {\bibfnamefont {G.}~\bibnamefont
  {Milione}}, \bibinfo {author} {\bibfnamefont {H.~I.}\ \bibnamefont {Sztul}},
  \bibinfo {author} {\bibfnamefont {D.~A.}\ \bibnamefont {Nolan}},\ and\
  \bibinfo {author} {\bibfnamefont {R.~R.}\ \bibnamefont {Alfano}},\ }\bibfield
   {title} {\bibinfo {title} {Higher-order poincar$\rm\acute{e}$ sphere, stokes
  parameters, and the angular momentum of light},\ }\href
  {https://doi.org/10.1103/PhysRevLett.107.053601} {\bibfield  {journal}
  {\bibinfo  {journal} {Phys. Rev. Lett.}\ }\textbf {\bibinfo {volume} {107}},\
  \bibinfo {pages} {053601} (\bibinfo {year} {2011})}\BibitemShut {NoStop}%
\bibitem [{\citenamefont {Naidoo}\ \emph {et~al.}(2016)\citenamefont {Naidoo},
  \citenamefont {Roux}, \citenamefont {Dudley}, \citenamefont {Litvin},
  \citenamefont {Piccirillo}, \citenamefont {Marrucci},\ and\ \citenamefont
  {Forbes}}]{Naidoo16}%
  \BibitemOpen
  \bibfield  {author} {\bibinfo {author} {\bibfnamefont {D.}~\bibnamefont
  {Naidoo}}, \bibinfo {author} {\bibfnamefont {F.~S.}\ \bibnamefont {Roux}},
  \bibinfo {author} {\bibfnamefont {A.}~\bibnamefont {Dudley}}, \bibinfo
  {author} {\bibfnamefont {I.}~\bibnamefont {Litvin}}, \bibinfo {author}
  {\bibfnamefont {B.}~\bibnamefont {Piccirillo}}, \bibinfo {author}
  {\bibfnamefont {L.}~\bibnamefont {Marrucci}},\ and\ \bibinfo {author}
  {\bibfnamefont {A.}~\bibnamefont {Forbes}},\ }\bibfield  {title} {\bibinfo
  {title} {Controlled generation of higher-order poincar$\rm\acute{e}$ sphere
  beams from a laser},\ }\href {https://doi.org/10.1038/NPHOTON.2016.37}
  {\bibfield  {journal} {\bibinfo  {journal} {Nat. Photon.}\ }\textbf {\bibinfo
  {volume} {10}},\ \bibinfo {pages} {327} (\bibinfo {year} {2016})}\BibitemShut
  {NoStop}%
\bibitem [{\citenamefont {Liu}\ \emph {et~al.}(2017)\citenamefont {Liu},
  \citenamefont {Liu}, \citenamefont {Ke}, \citenamefont {Liu}, \citenamefont
  {Shu}, \citenamefont {Luo},\ and\ \citenamefont {Wen}}]{Liu17}%
  \BibitemOpen
  \bibfield  {author} {\bibinfo {author} {\bibfnamefont {Z.}~\bibnamefont
  {Liu}}, \bibinfo {author} {\bibfnamefont {Y.}~\bibnamefont {Liu}}, \bibinfo
  {author} {\bibfnamefont {Y.}~\bibnamefont {Ke}}, \bibinfo {author}
  {\bibfnamefont {Y.}~\bibnamefont {Liu}}, \bibinfo {author} {\bibfnamefont
  {W.}~\bibnamefont {Shu}}, \bibinfo {author} {\bibfnamefont {H.}~\bibnamefont
  {Luo}},\ and\ \bibinfo {author} {\bibfnamefont {S.}~\bibnamefont {Wen}},\
  }\bibfield  {title} {\bibinfo {title} {Generation of arbitrary vector vortex
  beams on hybrid-order poincar$\rm\acute{e}$ sphere},\ }\href
  {https://doi.org/10.1364/PRJ.5.000015} {\bibfield  {journal} {\bibinfo
  {journal} {Photon. Res.}\ }\textbf {\bibinfo {volume} {5}},\ \bibinfo {pages}
  {15} (\bibinfo {year} {2017})}\BibitemShut {NoStop}%
\bibitem [{\citenamefont {Erhard}\ \emph {et~al.}(2018)\citenamefont {Erhard},
  \citenamefont {Fickler}, \citenamefont {Krenn},\ and\ \citenamefont
  {Zeilinger}}]{Erhard18}%
  \BibitemOpen
  \bibfield  {author} {\bibinfo {author} {\bibfnamefont {M.}~\bibnamefont
  {Erhard}}, \bibinfo {author} {\bibfnamefont {R.}~\bibnamefont {Fickler}},
  \bibinfo {author} {\bibfnamefont {M.}~\bibnamefont {Krenn}},\ and\ \bibinfo
  {author} {\bibfnamefont {A.}~\bibnamefont {Zeilinger}},\ }\bibfield  {title}
  {\bibinfo {title} {Twisted photons: new quantum perspectives in high
  dimensions},\ }\bibfield  {journal} {\bibinfo  {journal} {Light: Science \&
  Applications}\ }\textbf {\bibinfo {volume} {7}},\ \href
  {https://doi.org/10.1038/lsa.2017.146} {10.1038/lsa.2017.146} (\bibinfo
  {year} {2018})\BibitemShut {NoStop}%
\bibitem [{\citenamefont {Andrews}(2021)}]{Andrews21}%
  \BibitemOpen
  \bibfield  {author} {\bibinfo {author} {\bibfnamefont {D.~L.}\ \bibnamefont
  {Andrews}},\ }\bibfield  {title} {\bibinfo {title} {Symmetry and quantum
  features in optical vortices},\ }\href {https://doi.org/10.3390/sym.13081368}
  {\bibfield  {journal} {\bibinfo  {journal} {Symmetry}\ }\textbf {\bibinfo
  {volume} {13}},\ \bibinfo {pages} {1368} (\bibinfo {year}
  {2021})}\BibitemShut {NoStop}%
\bibitem [{\citenamefont {Angelsky}\ \emph {et~al.}(2021)\citenamefont
  {Angelsky}, \citenamefont {Bekshaev}, \citenamefont {Dragan}, \citenamefont
  {Maksimyak}, \citenamefont {Zenkova},\ and\ \citenamefont
  {Zheng}}]{Angelsky21}%
  \BibitemOpen
  \bibfield  {author} {\bibinfo {author} {\bibfnamefont {O.~V.}\ \bibnamefont
  {Angelsky}}, \bibinfo {author} {\bibfnamefont {A.~Y.}\ \bibnamefont
  {Bekshaev}}, \bibinfo {author} {\bibfnamefont {G.~S.}\ \bibnamefont
  {Dragan}}, \bibinfo {author} {\bibfnamefont {P.~P.}\ \bibnamefont
  {Maksimyak}}, \bibinfo {author} {\bibfnamefont {C.~Y.}\ \bibnamefont
  {Zenkova}},\ and\ \bibinfo {author} {\bibfnamefont {J.}~\bibnamefont
  {Zheng}},\ }\bibfield  {title} {\bibinfo {title} {Structured light control
  and diagnostics using optical crystals},\ }\href
  {https://doi.org/10.3389/fphy.2021.715045} {\bibfield  {journal} {\bibinfo
  {journal} {Front. Phys.}\ }\textbf {\bibinfo {volume} {9}},\ \bibinfo {pages}
  {715045} (\bibinfo {year} {2021})}\BibitemShut {NoStop}%
\end{thebibliography}%

\end{document}